\documentstyle[apb_cite]{mn}

\newcommand{\lta}{{\small\raisebox{-0.6ex}{$\,\stackrel
{\raisebox{-.2ex}{$\textstyle <$}}{\sim}\,$}}}

\begin{document}

\title[Fireball Models for Flares in AE~Aquarii]
{Fireball Models for Flares in AE~Aquarii}

\author[Pearson, Horne, Skidmore]
{K.\ J.\ Pearson\thanks{Send offprint requests to: 
kjp1@rouge.phys.lsu.edu}\thanks{Present address: Lousiana State University,
Department of Physics and Astronomy, Nicholson Hall, Baton Rouge, 
LA 70803-4001, USA},
Keith Horne, Warren Skidmore\thanks{Present address: Department of Physics and
Astronomy, University of California, Irvine, 4129 Frederick Reines Hall,
Irvine, CA 92697-4575, USA}
\\ School of Physics and Astronomy,
	University of St. Andrews, North Haugh, St Andrews KY16 9SS \\
}

\date{Accepted . Received ; 
in original form }

\maketitle

\begin{abstract} 

We examine the flaring behaviour of the cataclysmic variable AE~Aqr in the 
context of the `magnetic propeller' model for this system. The flares are
thought to arise from collisions between high density regions in the 
material expelled from the system after interaction with the
rapidly rotating magnetosphere of the white dwarf. We calculate the 
first quantitative models for the flaring and calculate the time-dependent 
emergent optical spectra from the resulting hot, expanding ball of gas.
We compare the results under different assumptions to
observations and derive values for the mass, lengthscale and temperature 
of the material involved in the flare. We see that the fits suggest that the
secondary star in this system has Population II composition.

\end{abstract} 


\section{Introduction}
\label{sec:intro}

AE~Aqr is an unusual cataclysmic variable star exhibiting
bizarre phenomena that can now be interpreted in the context of a
magnetic propeller that throws gas out of the binary system
\cite{wynn97}. The magnetic propeller efficiently extracts energy and angular
momentum from the white dwarf and transports it via magnetic fields to gas
stream material which it ejects from the binary system.
Detailed understanding of the magnetic propeller effects
may therefore help us to unlock some of the secrets
of magnetic viscosity in accretion flows.

In the AE~Aqr binary system, a slightly evolved K star \cite{welsh95}
that overflows its Roche lobe is locked in a 9.88 hour orbit
with a rapidly spinning magnetized white dwarf.
Coherent oscillations in optical \cite{patterson79}, ultraviolet 
\cite{eracleous94}, and X-ray \cite{patterson80} lightcurves
reveal the white dwarf's 33s spin period. 
The oscillation has two unequal peaks per spin cycle,
consistent with broad hotspots on opposite sides of the white dwarf
$\sim 15^\circ$ above and below the equator \cite{eracleous94}.
These oscillations are strongest in the ultraviolet,
where their spectra show a blue continuum with broad Ly~$\alpha$ absorption
consistent with a white dwarf ($\log{g}\approx8$) atmosphere 
with $T\sim 3\times10^4~{\rm K}$ \cite{eracleous94}.
The simplest interpretation is accretion heating
near the poles of a magnetic dipole field tipped almost perpendicular
to the rotation axis.

An 11-year study of the optical oscillation period \cite{dejager94} shows
that the white dwarf is spinning down at an alarming rate.
Something extracts rotational energy from the white dwarf
at a rate $I\omega\dot{\omega} \sim 60 \nu L_\nu$ 
ie. some 60 times the luminosity of the system.
We now believe that to be a magnetic propeller. This model was
first proposed by \scite{wynn95} and \scite{eracleous96} and expanded upon
in \scite{wynn97} with comparison of observed and modelled tomograms.
Further work on the flaring region was reported
in \scite{welsh98} and \scite{horne99}.

The gas stream emerging from the companion star through the L1 nozzle
encounters a rapidly spinning magnetosphere.
The rapid spin makes the effective ram pressure so high
that only a low-density fringe of material becomes threaded
onto field lines.
Most of the stream material remains diamagnetic, and is
dragged toward co-rotation with the magnetosphere.
As this occurs outside the co-rotation radius, this magnetic drag
propels material forward, boosting its velocity up to and beyond
escape velocity.
The material emerges from the magnetosphere and sails out of the
binary system.
This process efficiently extracts energy and angular momentum
from the white dwarf, transferring it via the long-range magnetic field
to the stream material, which is expelled from the system.
The ejected outflow consists of a broad equatorial fan of material
launched over a range of azimuths on the side away from the K star.

The material stripped from the gas stream and threaded by the field
lines has a different fate, one which we believe gives rise to the
radio and X-ray emission.
This material co-rotates with the magnetosphere while accelerating
along field lines either toward or away from the white dwarf
under the influences of gravity and centrifugal forces.
The small fraction of the total mass transfer that leaks below the co-rotation 
radius at $\sim5~R_{\rm wd}$
accretes down field lines producing the surface hotspots 
responsible for the 33s oscillations.
Particles outside co-rotation remain trapped long enough to
accelerate up to relativistic energies through magnetic pumping,
eventually reaching a sufficient energy density to break away
from the magnetosphere \cite{kuijpers97}.
The resulting ejection of balls of relativistic magnetized plasma
is thought to give rise to the flaring radio emission 
\cite{bastian88a,bastian88b}.

This paper addresses the optical and ultraviolet variability
seen in AE~Aqr.
In many studies the lightcurves exhibit dramatic flares,
with 1-10 minute rise and fall times 
\cite{patterson79,paradijs89,bruch91,welsh93}.
The flares seem to come in clusters or avalanches of many super-imposed 
individual flares separated by quiet intervals of gradually
declining line and continuum emission \cite{eracleous96,patterson79}.
These quiet and flaring states typically last a few hours.
Power spectra computed from the lightcurves have a power-law form,
with larger amplitudes on longer timescales.
\scite{elsworth82} found that the index in
$A\propto\nu^{\alpha}$ was -1. That 
is, that the amplitude of the flare or flicker varies inversely as the 
frequency of its occurence.
\scite{bruch92} examined several datasets and found values for $\alpha$
in the range $-0.71$--$-1.64$. 
Such power-law spectra are often associated with physical processes 
involving self-organized criticality, for example earthquakes,
snow or sandpile avalanches \cite{bak96}.
Similar red noise power spectra are seen in 
active galaxies, X-ray binaries, and other cataclysmic variables,
and is therefore regarded as characteristic of accreting sources in general.
However, flickering in other cataclysmic variables
typically has an amplitude of 5-20\% \cite{bruch92}, contrasting with
factors of several in AE~Aqr.
If the mechanism is the same, then it must be weaker or
dramatically diluted in other systems.

The optical and ultraviolet spectra of the AE~Aqr flares 
are not understood at present except in the most general terms.
The lines and continua rise and fall together, with little change
in the equivalent widths or ratios of the emission lines \cite{eracleous96}.
This suggests that the flares represent changes in the amount
of material involved more than changes in physical conditions.
The Balmer emission lines decay somewhat more slowly than the 
optical continuum -- perhaps revealing a recombination time delay 
\cite{welsh98}.
Ultraviolet spectra from HST reveal a wide range of lines representing
a diverse mix of ionization states and densities.
\scite{eracleous96} conclude:
\begin{quote}
``Based on the critical densities of the observed semiforbidden lines we
suggest that in a large fraction of the line-emitting gas the density is in 
the range $n\sim10^{9}$--$10^{11}~{\rm cm}^{-3}$. It is likely that denser 
regions also exist."
\end{quote}
 thus setting a lower limit on the density range in the
flaring region.
Such spectra suggest shocks. The C{\sc \small IV} emission is unusually weak, 
suggesting non-solar abundances.
For example, carbon depletion may occur if CNO-processed material
is being transferred from the secondary star, which is an evolved star
being whittled down by Roche lobe overflow.
So far no quantitative fits to the spectra have been
achieved either by shock or photo-ionization models. IUE observations by
\scite{jameson80} derive an emission measure ($V_{\rm HII}N_{\rm e}^{2}$) 
from Ly$\alpha$ of $10^{61}~\mbox{m}^{-3}$. UBVRI colour photometry of flares 
by \scite{beskrovnaya96} give colour indices close to those for a 
blackbody in the $15~000$--$20~000$K range with an emitting area
$\sim10^{16}~\mbox{m}^{2}$.

What mechanism triggers these dramatic optical and ultraviolet flares?  
Clues come from multi-wavelength co-variability and 
orbital kinematics.
Simultaneous VLA and optical observations show that the radio 
flux variations occur on similar timescales but are not correlated
with the optical and ultraviolet flares,
which therefore require a different mechanism \cite{abada-simon95}.
It was proposed that the flares represent modulations
of the accretion rate onto the white dwarf, so that they should be
correlated with X-ray variability.
Some correlation was found, but the correlation is not high.
However, HST observations discard this model, because
the ultraviolet oscillation amplitude is unmoved by
transitions between the quiet and flaring states \cite{eracleous96}.
This disconnects the origins of the oscillations and flares,
and the oscillations arise from accretion onto the white dwarf,
so the flares must arise elsewhere.

Further clues come from emission line kinematics.
The emission line profiles may be roughly described as
broad Gaussians with widths $\sim1~000$~km~s$^{-1}$,
though they often exhibit kinks and sometimes multiple peaks.
Detailed study of the Balmer lines \cite{welsh98} indicates that the
new light appearing during a flare can have emission lines
shifted from the line centroid and somewhat narrower,
$\sim300$~km~s$^{-1}$.
Individual flares therefore occupy only a subset of the entire
emission-line region.

The emission line centroid velocities vary sinusoidally
with orbital phase, with semi-amplitudes $\sim200$~km~s$^{-1}$
and maximum redshift near phase $\sim$0.8 \cite{welsh98}.
These unusual orbital kinematics are shared by both 
ultraviolet and optical emission lines.
The implication is that the flares arise from gas
moving with a $\sim200$~km~s$^{-1}$ velocity vector
that rotates with the binary and points away the observer at phase 0.8.
This is hard to understand in the standard model
of a cataclysmic variable star, though many cataclysmic variables
show similar anomalous emission-line kinematics \cite{thorstensen91}.
Kepler velocities in an AE~Aqr accretion disc 
would be $>600$~km~s$^{-1}$ (though we believe no disc to be present.)
The gas stream has a similar direction but its velocity is
$\sim1~000$~km~s$^{-1}$.
A success of the magnetic propeller models is its ability to
account for the anomalous emission-line kinematics.
The correct velocity amplitude and direction occurs
in the exit fan just outside the Roche lobe of the white dwarf.
But the question remains of why the flares are ignited here, several hours
after the gas slips silently through the magnetosphere.

\begin{figure*}
\begin{minipage}{17cm}
\vspace{7.0cm}
\includegraphics{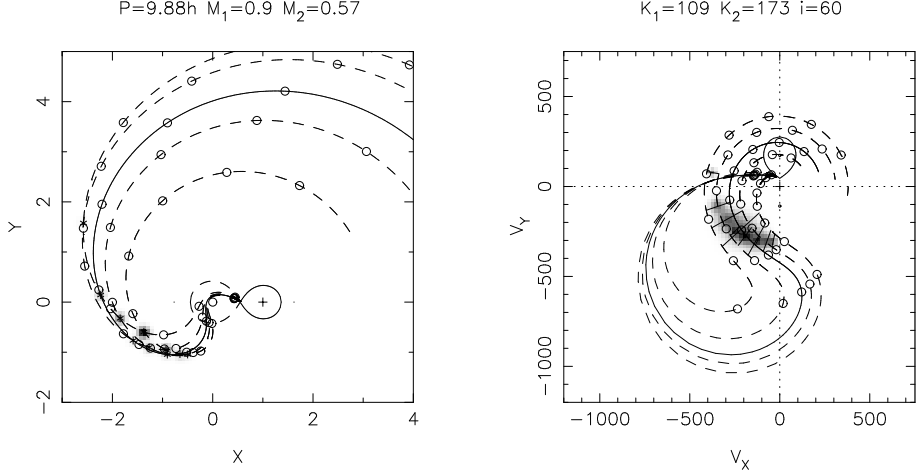}
\caption{Trajectories of diamagnetic blobs passing though the AE~Aquarii 
system with the corresponding Doppler tomogram. The open circles mark the 
time of flight in units of $0.1~P_{\rm orb}$. Asterisks and grey patches
mark the locations where lower density blobs overtake and collide with more 
compact blobs, producing fireballs. This occurs in the lower left quadrant
at the location corresponding to the observed emission lines.}
\label{fig:trajdop}
\end{minipage}
\end{figure*}

The key insight which solved this puzzle was the realization
that the magnetic propeller acts as a blob sorter.
More compact, denser, diamagnetic blobs are less affected by magnetic drag.
They punch deeper into the magnetosphere and
emerge at a larger azimuth with a smaller terminal velocity.
These compact blobs can therefore be overtaken by `fluffier' blobs
ejected with a larger terminal velocity in the same direction,
having left the companion star somewhat later
but spent less time in the magnetosphere \cite{wynn97}.
The result is a collision between two gas blobs, which can give rise
to shocks and flares.
Calculations of the trajectories of magnetically propelled diamagnetic
blobs with different drag coefficients indicate that they
cross in an arc-shaped region of the exit stream, 
in just the right place to account for the orbital kinematics of
the emission lines \cite{welsh98,horne99}. Figure~\ref{fig:trajdop} shows
how these trajectories map to a locus
of points in the lower left quadrant of the Doppler map that is 
otherwise difficult to populate.

There remains the problem of developing a quantitative understanding
of the unusual emission-line spectra of the blobs.
We have two aims for this paper: the principle aim is to see if the observed 
peak optical flare spectrum can be reproduced by conditions appropriate to the 
aftermath of a collision between blobs.
The secondary objective is to develop as far as possible an understanding 
of the mechanisms by which the observed spectra are formed and 
evolve. We have approached both goals in the spirit of
attempting to find the simplest models that explain the observations.

In section~\ref{sec:models}
we outline the basic assumptions used in our models. We go on in
section~\ref{sec:radtrans} to consider the radiative transfer problem and 
calculate analytic expressions for the behaviour of the optical flare spectra 
and lightcurves and outline some numerical considerations .
In section~\ref{sec:results} we present the results of numerical 
simulations of the flare behaviour, both lightcurves and spectra, 
and consider further the limits of applicability of our method. We discuss 
the 
interpretation of our results in section~\ref{sec:discuss} in terms 
of the fireball model. Finally we summarise our results in 
section~\ref{sec:summary}.

\section{Fireball Models}
\label{sec:models}

\subsection{Rough Estimates}
\label{sec:estimate}

We derive rough estimates for the physical parameters associated with
the colliding blobs by 
using the typical rise time $t\sim300~{\rm s}$ and
an observed optical flux $f_{\nu}\sim50$~mJy with a  
closing velocity $V\sim300~{\rm km~s}^{-1}$ for the blobs.
We estimate a mass transfer rate
\begin{equation}
\dot{M}\sim\frac{2 \nu L_{\nu}}{V^{2}}=\frac{2\nu f_{\nu} d^2}{V^{2}} \approx
10^{14}~{\rm kg~s}^{-1}
\end{equation} 
using a distance of 100pc. This compares to a value from the standard 
evolutionary equation
\begin{equation}
-\dot{M_2}\approx6\times10^{-10}\left(\frac{P_{\rm h}}{3}\right)^{\frac{5}{3}}
 M_\odot {\rm y}^{-1} = 3\times10^{14}~{\rm kg}~{\rm s}^{-1}
\label{eqn:mdot}
\end{equation}
\cite{FKR}. The mean density of the overflowing gas stream
in a CV can be calculated from
\begin{equation}
\bar{\rho}=\frac{\dot{M}}{QV}
\end{equation}
where the stream cross-section
\begin{equation}
Q\approx2.4\times10^{13}\left(\frac{T_{\rm s}}{10^4~{\rm K}}\right)
{P^{2}_{\rm orb}({\rm h})~{\rm m}^{2}}
\end{equation}
has been derived by several authors 
\cite{papaloizou75,meyer83,hameury86,ritter88,kovetz88,sarna90,warner95}. 
With a secondary temperature 
$T_{\rm s}=4~500~{\rm K}$ \cite{skidmore02}, we have a mean density,
\begin{equation}
\bar{\rho} \approx 
3.2\times10^{-8}~\left(\frac{V}{300~{\rm km~s}^{-1}}\right)^{-1}
{\rm kg~m}^{-3}
\end{equation}
ie.
\begin{equation}
\bar{n}\approx\frac{\bar{\rho}}{\mu m_{\rm H}} = 3.2\times10^{19}~{\rm m}^{-3}
\end{equation}
where $\mu\approx0.6$ is the mean molecular weight appropriate for a fully
ionized gas with solar composition.
This is consistent with the lower limits from HST uv observations discussed 
earlier. It must be remembered that this is the {\it mean} density in
a smooth stream. A stream composed of discrete blobs of material would have 
range of densities about this value.  

We estimate the total mass involved in the collision to be
\begin{equation}
M \sim \dot{M} t \sim3\times10^{16}~{\rm kg}.
\end{equation}
The typical pre-collision lengthscale $a$ of the problem is given by
\begin{equation}
a\sim V t \sim 9\times10^{7}~{\rm m}.
\end{equation}
Following the collision we might expect a somewhat smaller lengthscale 
eg.$\sim5\times10^{7}~{\rm m}$. 
The above density then suggests an emission measure 
$Qa\bar{n}^2\approx10^{63}~{\rm m}^{-3}$ consistent with the above IUE 
measurements for an optically thick Ly$\alpha$ line.

The initial post collision temperature $T$ following a strong shock
\begin{equation}
T\sim \frac{3}{16}\frac{\mu m_{\rm H} V^{2}}{k} \approx1.2\times10^{6}~{\rm K}.
\end{equation}
Finally we have the energy involved in the collision and maximum possible 
flare energy $E$
\begin{equation}
E \sim \frac{M V^{2}}{8} \approx 3\times10^{26}~{\rm J}.
\end{equation}
A summary of typical values is provided in Table~\ref{tab:scaleval}.

\begin{table}
\begin{tabular}{rrl} \hline
Quantity & \multicolumn{2}{c}{Value} \\ \hline 
Observed Flux $f_{\nu}$ & $50$ & mJy\\
Closing Velocity $V$ & $300$ & $\rm{km~s}^{-1}$\\
Flare Risetime $t_{\rm rise}$ & 300 & s \\
Mass Transfer Rate $\dot{M}$ & $10^{14}$ & ${\rm kg~s}^{-1}$ \\
Fireball Mass $M$ & $3\times10^{16}$ & kg \\
Pre-Collision Lengthscale $a$ & $9\times10^{7}$ & m \\
Initial Temperature $T$ & $10^{6}$ & K \\
Total Energy $E$ & $3\times10^{26}$ & J \\
Typical Density $\rho$ & $3\times10^{-8}$ & ${\rm kg~m}^{-3}$ \\
Number Density $n$ & $3\times10^{19}$ & ${\rm m}^{-3}$ \\
Column Density $N$ & $3\times10^{27}$ & ${\rm m}^{-2}$ \\ \hline 
\end{tabular}
\caption{Estimates of typical values for flare quantities.}
\label{tab:scaleval}
\end{table}

\subsection{Initial Conditions}

We envision an expanding fireball emerging from the aftermath of
a collision between two masses $m_1+m_2=M$.
In the centre of mass frame, these masses
approach with initial velocities satisfying $m_1 v_1 = m_2 v_2$.
The initial kinetic energy in the centre of mass frame is
\begin{equation}
E = \frac{ m_1 v_1^2 + m_2 v_2^2 }{2} = \frac{ \displaystyle M V^2 q}
	{ \displaystyle 2 \left( 1 + q \right)^2 }
\end{equation}
where $V = v_1 + v_2$ is the (frame-independent) relative velocity
and $q = m_1/m_2 (= v_2/v_1)$ is the mass ratio.
Note that for a given $M$ and $V$ a range of energies is possible,
ranging from zero for very extreme mass ratios up to
a maximum of $E_0 = MV^2/8$ for equal masses.

As the collision progresses, some of the energy $E$ is converted
to thermal energy through dissipation in shocks.
The initial dissipative phase lasts a short time
\begin{equation}
t_{\rm i} \sim \frac{a_{\rm i}}{V} = 300 
\left(\frac{a_{\rm i}}{10^{8}~\mbox{m}}\right)
\left(\frac{V}{300~\mbox{km~s}^{-1}}\right)^{-1}~\mbox{s}. 
\end{equation}

The ratio $E/M$ of energy per mass sets the
initial temperature $T_{\rm i}$ and sound speed $c_{\rm s,i}$ through
\begin{equation}
\frac{E}{M} \approx \frac{\gamma k T_{\rm i}}{\mu m_H} \approx{c_{\rm s,i}}^2 ,
\label{initT}
\end{equation}
where a specific heat ratio $\gamma = 5/3$
is appropriate for fully ionized atomic gas.
For $V\sim300$~km~s$^{-1}$, $T_{\rm i} \sim 10^{6}~{\rm K}$, 
so that the initial ball of hot gas is atomic and completely ionized.

Since $V \sim c_{\rm s,i}$, sound waves have time to cross the
initial fireball during the collision time $t_{\rm i}$, and we therefore 
expect a roughly uniform temperature profile inside $a_{\rm i}$.

\subsection{Expansion}

With nothing to hold it back, the hot ball of gas expands
at the initial sound speed, launching a fireball. We adopt a uniform, 
spherically symmetric, Hubble-like
expansion $V=Hr_{0}=Hr/\beta$, in which the Eulerian radial coordinate $r$
 of a gas element is given 
in terms of its initial position $r_0$ and time $t$ by
\begin{equation}
r(r_{0},t)=r_{0}+v(r_{0})t=r_{0}+Hr_{0}t=r_{0}(1+Ht)\equiv r_{0}\beta.
\label{expfac}
\end{equation}
This defines an expansion factor $\beta$ which we can use as a dimensionless 
time parameter. The `Hubble' constant is set by the initial conditions
$H a_{\rm i} \approx v(a_{\rm i}) \approx c_{\rm s,i}$.

Uniform free expansion is a suitable approximation because the flow becomes
supersonic as it expands and cools.
If we can determine parameters at some time $t=0$ ie. $\beta=1$ for the
lengthscale ($a_0$), temperature ($T_0$) and mass ($M$) we can derive the time
evolution as follows. We adopt a Gaussian density profile
\begin{equation}
\rho (r,t) = \rho_{0} \beta^{-3} e^{-\eta^{2}}
\label{denprof}
\end{equation}
where
\begin{equation}
\rho_{0} = \frac{M}{\left( \pi a_{0}^{2} \right)^{\frac{3}{2}}}
\end{equation}
is the initial central density and
\begin{equation}
\eta\equiv\frac{r_{0}}{a_{0}}=\frac{r}{a}
\end{equation}
is a dimensionless radius coordinate, the radius $r$ scaled to the lengthscale
$a=\beta a_{0}$.

Although this Gaussian density profile is only a guess, it is motivated by the
Gaussian shapes of
observed velocity profiles, and by the thought that the initial
thermal velocity distribution in the hot gas will map into the
density profile of the expanding fireball because the
fast particles travel farther than slow ones. 

\subsection{Cooling}

We consider three cooling schemes: adiabatic, isothermal and radiative.
The adiabatic fireball cools purely as a result of its expansion
and corresponds to a situation where the radiative and 
recombination cooling rates are negligible. In contrast, the isothermal model 
maintains a fixed temperature throughout its evolution. A truly isothermal 
fireball would require a finely balanced energy source to 
counteract the expansion cooling. However, it may an appropriate 
approximation to a 
situation where a photospheric region dominates the emission and presents a  
fixed effective temperature to the observer as a result of the stong dependence of opacity
on temperature. The radiative
models cool adiabatically throughout and as a result of radiation from 
a thin zone near the photosphere which we model by immediately dropping
the temperature to $T_{\rm p}\sim10^{4}~{\rm K}$ at this surface. 

\subsubsection{Adiabatic Cooling}

During adiabatic expansion, $P\propto\rho^\gamma$
and $T \propto P/\rho \propto \rho^{\gamma-1}$.
If we consider the evolution of a given gas element then, with 
$\rho\propto\beta^{-3}$, the temperature decreases with time as
\begin{equation}
	T(r,t) = T_0(r_{0}) \beta^{-3(\gamma-1)}.
\end{equation}
We see that an initially uniform temperature distribution remains
uniform and so, for this case, we can write, more simply,
\begin{equation}
	T(r,t) = T_0 \beta^{-3(\gamma-1)}.
\label{temptdep}
\end{equation}
With $\gamma=5/3$ for a monatomic gas, $T\propto \beta^{-2}$, and so
the sound speed $c_{\rm s} \propto T^{1/2} \propto \beta^{-1}$.
The sound crossing time $a/c_{\rm s} \propto \beta^2$.
The fireball becomes almost immediately supersonic.
We can derive the behaviour of the sonic point from the definition
that it is the position where the expansion rate matches the sound speed. Thus,
\begin{equation}
1 \equiv \frac{v(r_{\rm s},t)}{c_{\rm s}(t)} = 
\frac{H r_{{\rm s},0}}{c_{\rm s}(t)} = \frac{H r_{\rm s}}{\beta c_{\rm s}(t)}
= \frac{H r_{\rm s}}{c_{{\rm s,}0}}.
\end{equation}
Assuming a uniform temperature profile and $\gamma=5/3$, this gives
\begin{equation}
r_{\rm s}  = \frac{c_{{\rm s,}0}}{H} = a_{0}.
\end{equation}
Hence, we can see that the sonic point remains fixed in space at $a_0$.
The initial density structure emerging through the sonic
surface at radius $a_0$ is subsequently `frozen in' by the rapid expansion.

\subsubsection{Radiative Cooling Timescales}

The time required for the fireball to radiate away its internal energy 
depends crucially on the temperature of the photosphere (assuming that 
the fireball is optically thick). For a uniform temperature distribution 
the timescale is very short: 
\begin{eqnarray}
\frac{U}{\dot{U}} & \sim & \frac{M k}{m_{\rm H} \sigma R^2 T^3} \\ 
                  & = & 1.5 \left( \frac{M}{10^{17}~{\rm kg}}\right) 
\left(\frac{R}{10^{8}~{\rm m}}\right)^{-2} 
\left(\frac{T}{10^{4}~{\rm K}}\right) ^{-3} {\rm s}. 
\end{eqnarray}
We would thus expect the region from which the radiation 
escapes to cool rapidly until the region becomes optically thin and the
cooling efficiency drops.

If the photospheric temperature $T_{\rm p}$ is significantly cooler than 
the optically 
thick region, the radiative timescale
\begin{eqnarray}
\frac{U}{\dot{U}} & \sim & \frac{M k T}{m_{\rm H} \sigma R^2 T_{\rm p}^4} \\ 
                  & = & 150\left( \frac{M}{10^{17}~{\rm kg}}\right) 
\left(\frac{R}{10^{8}~{\rm m}}\right)^{-2} \nonumber \\
     & ~  & ~~~~~~~~~~~~~~~~~~~~\times \left(\frac{T}{10^{6}~{\rm K}} \right)
\left(\frac{T_{\rm p}}{10^{4}~{\rm K}}\right) ^{-4} {\rm s}  
\end{eqnarray}
can become comparable to the evolutionary timescale of the fireball.

\subsubsection{Radiative Cooling Front}

Following the exact behaviour of radiative cooling and heating is a complex
process requiring us to follow the behaviour of the internal radiation
field of the fireball through the optically thick to optically thin transition.
Such detailed modelling is beyond the scope of the initial investigations 
presented here. We note that studies of cooling from optically thin plasmas 
\cite{lyndenbell01,dalgarno72} show that
there is a significant reduction in efficiency due to hydrogen 
recombination and the consequent loss of free electrons below 
approximately $10^4$~K.
Since the photosphere is the region where the optical depth for
escaping radiation is unity, this suggests that the photosphere may adopt a 
roughly constant effective temperature $T_{\rm p}\sim 10^{4}~\mbox{K}$.
The relatively dense material in the fireball will be able to cool rapidly down
to this temperature before the cooling rate stalls.
We approximate a radiatively cooling fireball by maintaining a 
hot core region which cools adiabatically as it expands. Once gas elements 
cross the boundary to this region, we assume them to cool immediately to 
the effective temperature of the photosphere and thereafter adiabatically. We 
derive the radius of the photospheric surface below, using the blackbody 
luminosity and the  thermal energy of the central region. 

\begin{figure}
\vspace{6.25cm}
\includegraphics{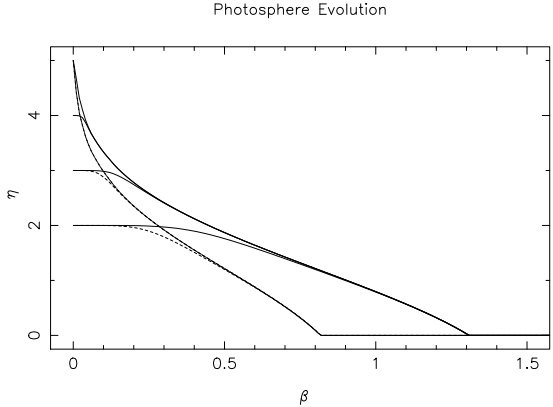}
\caption{The evolution of the temperature boundary for the radiative model with
$M=6.8\times10^{16}~{\rm kg}$, $a_{0}=9.6\times10^7~{\rm m}$ and 
$T_0=18~000~{\rm K}$. The solid line is for $T_{\rm p}=10~000$~K and 
the dashed line for $T_{\rm p}=18~000$~K. In each case we have initial values
for $\eta_{\rm p}=2,3,4,5$.}  
\label{fig:tempbnd}
\end{figure}

\begin{figure}
\vspace{6.25cm}
\includegraphics{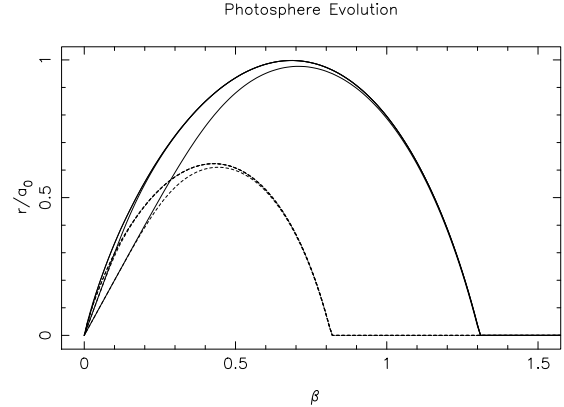}
\caption{The evolution of the temperature boundary in inertial coordinates
for the radiative model with
$M=6.8\times10^{16}~{\rm kg}$, $a_{0}=9.6\times10^7~{\rm m}$ and 
$T_0=18~000~{\rm K}$. The solid line is for $T_{\rm p}=10~000$~K and 
the dashed line for $T_{\rm p}=18~000$~K. In each case we have initial values
for $\eta_{\rm p}=2,3,4,5$.} 
\label{fig:abstempbnd}
\end{figure}

To find the photospheric radius $r_{\rm p}=a_{0} \beta \eta_{\rm p}$ in our 
radiative model we 
calculate the total thermal energy in a sphere of radius $r_{\rm p}$
\begin{eqnarray}
U & =& \int_{0}^{r_{\rm p}} \frac{3}{2} \frac{\rho k T}{\mu m_{\rm H}} 
4 \pi r^2 dr\\
& = & \frac{6 \pi k}{\mu m_{\rm H}} \frac{T_{0}a_{0}^{3} \rho_{{\rm c,}0}}
{\beta^{3(\gamma-1)}}
\int_{0}^{\eta_{\rm p}}  e^{-\eta^2} 
\eta^{2} d\eta.
\end{eqnarray}
Hence, differentiating,
\begin{equation}
\frac{ \partial U}{\partial \eta_{\rm p}} =
\frac{6 \pi k T_0 a_0^3 \rho_{{\rm c},0}}{\mu m_{\rm H} \beta^{3(\gamma-1)}}
\eta_{\rm p}^2  
e^{- \eta_{\rm p}^2}. 
\end{equation}
Radiative cooling at fixed effective temperature $T_{\rm p}$ gives
\begin{equation}
\frac{\partial U}{\partial \beta} = - 4 \pi H a_{0}^{2} \beta^{2}
\eta_{\rm p}^{2} \sigma T_{\rm p}^{4}
\end{equation}
and so, using
\begin{equation}
\frac{\partial \eta_{\rm p}}{\partial \beta} \equiv 
\left(\frac{\partial U}{\partial \beta}\right)
\left(\frac{\partial U}{\partial \eta_{\rm p}}\right)^{-1}, \\
\end{equation}
we have the differential equation for the core radius as a function of 
time
\begin{equation}
\frac{\partial \eta_{\rm p}}{\partial \beta} 
e^{-\eta_{\rm p}^{2}} 
= - C \beta^{3\gamma-1} ,
\label{eqn:cordiff}
\end{equation}
where
\begin{equation}
C =  \left( \frac{2 \mu m_{\rm H}\sigma}{3k} \right) 
\left( \frac{H T_{\rm p}^{4}}{\rho_{{\rm c,}0} T_0 a_0} \right). 
\end{equation}
We solve equation (\ref{eqn:cordiff}) numerically and plot the behaviour for 
typical parameters in Fig.~\ref{fig:tempbnd}. We see how, in
the Lagrangian coordinate $\eta$,  the boundary migrates inward continually. 
In 
Eulerian coordinates, plotted in Fig.~\ref{fig:abstempbnd}, the boundary 
is initially advected outwards with the flow. When the inward migration 
exceeds the expansion rate, the photosphere turns around and begins to 
collapse. In both plots the 
initial conditions for $\eta_{\rm p}$ have relatively little impact on the 
behaviour at late times. Models with $\eta_{\rm p}\ga4$ initially are 
indistinguishable as 
they rapidly converge on the same curve. Lower initial values for 
$\eta_{\rm p}$ also converge, albeit more slowly, on the same evolution.


\begin{figure}
\vspace{6.25cm}
\includegraphics{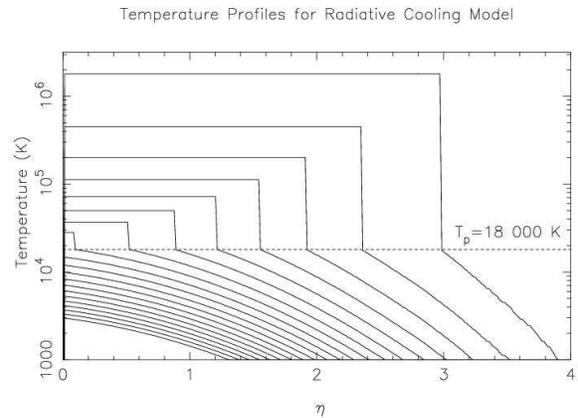}
\caption{The temperature profile for a radiative fireball with
$M=6.8\times10^{16}~{\rm kg}$, $a_{0}=9.6\times10^7~{\rm m}$, 
$T_0=18~000~{\rm K}$ and $T_{\rm p}=18~000~{\rm K}$. Time runs from 
$\beta=0.1$ for the hottest profile to
$\beta=2.0$ for the coolest in steps of $\beta=0.1$}
\label{fig:tprof}
\end{figure}

We plot the temperature profiles at different times for the radiative model
in Fig.~\ref{fig:tprof}. The profiles show how the material outside the
core region cools very rapidly with distance away from the boundary. As a 
result, the shell of material with significant temperature outside the
core is very thin. For computational convenience we set a minimum 
temperature for any of the gas at $1~000$~K.

\subsection{Ionization Structure}

In the LTE approximation, the density and temperature
determine the ionization state of the gas at each point in space and time
through the solution of a network of Saha equations.
Atomic level populations are similarly determined 
through Boltzmann factors and partition functions.
For purely adiabatic cooling, the temperature retains its initial
uniform spatial profile, but decreases with time.
The LTE ionization is therefore higher in the outer low-density
regions which also move fastest.
Thus in the LTE model high ionization emission lines 
are predicted to have broader velocity profiles.

\begin{figure}
\vspace{6.25cm}
\includegraphics{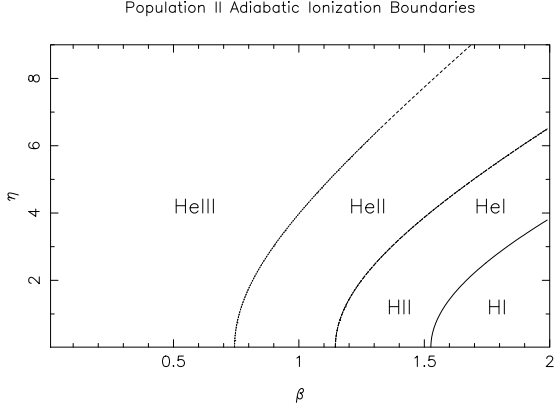}
\caption{Boundary between dominant species of hydrogen (solid) and helium 
(dashed) as a function of time for an adiabatic Population II 
model.}
\label{fig:adihionbnd}
\end{figure}

\begin{figure}
\vspace{6.25cm}
\includegraphics{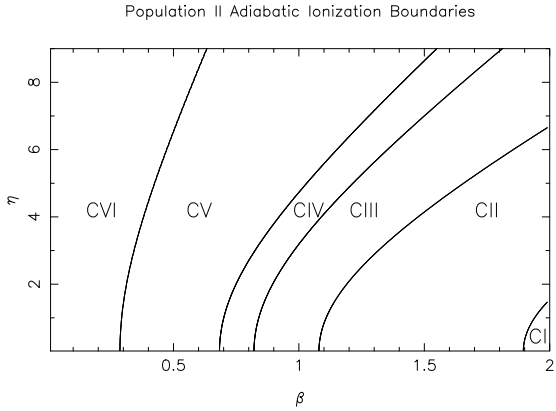}
\caption{Boundary between dominant species of carbon as a function of time for
an adiabatic Population II model.}
\label{fig:adicionbnd}
\end{figure}

\begin{figure}
\vspace{6.25cm}
\includegraphics{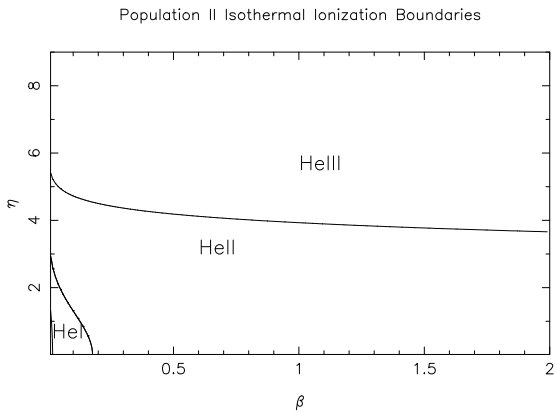}
\caption{Boundary between dominant species of helium 
as a function of time for an isothermal Population II model. Hydrogen
is fully ionised throughout the calculation.}
\label{fig:isohionbnd}
\end{figure}

\begin{figure}
\vspace{6.25cm}
\includegraphics{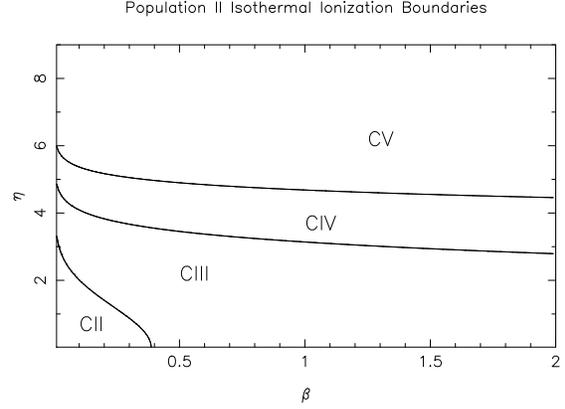}
\caption{Boundary between dominant species of carbon as a function of time for
an isothermal Population II model.}
\label{fig:isocionbnd}
\end{figure}

Once we have determined the evolution of $T$ and $\rho$ with time, 
we can follow
the evolution of the ionization structure and determine the boundaries 
between different ionization states of various species. We do so for hydrogen 
and helium 
for an adiabatic Population II model in Fig.~\ref{fig:adihionbnd} and 
for carbon in Fig.~\ref{fig:adicionbnd}. We plot boundaries for the same 
species for an isothermal Population II model in Figs.~\ref{fig:isohionbnd} 
and \ref{fig:isocionbnd}. In both cases we use $M=10^{18}~{\rm kg}$, 
$T_{0}=18~000~{\rm K}$ and $a_{0}=10^{8}~{\rm m}$.



At a given time, the various ionic species form onion layers of increasing
ionization on moving outwards from the fireball centre. The different 
temperature evolutions significantly alter the ionization evolution. The
adiabatic temperature evolution causes the ionization states to alter rapidly
throughout the structures when the fireball passes the appropriate
critical temperatures. The isothermal evolutions show the gentler dependance of
ionization on density. The spatial structure remains roughly constant and only
evolves slowly as the density drops.

\subsection{Ionization Freezing}
\label{sec:ionfreeze}

 For the initial models presented here we have
assumed that the ionization structure produced at any point
is given by LTE; an assumption we justify in Appendix~\ref{sec:ltevalidity}.
 In reality, of course, once the LTE boundary is crossed,
the ionization will drop as the net recombination rate gradually ekes away at 
the ions and we make a transition towards coronal equilibrium. However, 
since we know the time-evolution of density and temperature for any region 
outside of the LTE limiting radius, we can derive the time and conditions
when it crossed this LTE boundary. In principle then, we can follow the non-LTE
evolution of the gas through this outer region and beyond the ion-electron
equilibrium radius to the non-equilibrium ionization states present in the 
outer fringe.

In a purely LTE fireball we expect high ionization states to
occur in the outer regions as a result of the lower density. With the 
more detailed prescription above, the higher ionization would result from the 
higher temperature of
the fireball when the region in question crossed the LTE boundary. In practice,
we expect that this correction may have little effect on the optical spectra
because the emission is dominated by the higher density inner regions of the
fireball. This effect may well become important, however, when we come
to consider the ultraviolet spectra where high ionization lines are present; 
and which we hope to consider in a future paper. An alternative explanation 
for the ultraviolet behaviour may be that there is a hot, low density outer
fringe to the fireballs. These conditions may occur in regions that have 
always been optically thin and inefficient at cooling radiatively. These 
lines may then arise from a transition
region similar to that between the solar chromosphere and corona. 

\section{Radiative Transfer}
\label{sec:radtrans}

The radiative transfer equation has a formal solution
\begin{equation}
	I = \int S e^{-\tau} d\tau , \label{radtrans}
\end{equation}
where $I$ is the intensity of the emerging radiation,
$S$ is the source function, and $\tau$ is the optical depth
measured along the line of sight from the observer.
This integral sums contributions $S d\tau$ to the radiation intensity,
attenuating each by the factor $e^{-\tau}$ because 
it has to pass through optical depth $\tau$ to reach the
observer.

Since we have assumed LTE, the source function is the Planck function,
and opacities both for lines and continuum are also
known once the velocity, temperature, and density profiles
and element abundances are specified.
The integral can therefore be evaluated numerically, either in this form or 
more  quickly by using Sobolev resonant surface approximations. 

The above line integral gives the intensity
$I(y)$ for lines of sight with different impact parameters $y$.
We let $x$ measure distance from the fireball centre toward the
observer, and $y$ the distance perpendicular to the line of sight.
The fireball flux, obtained by summing intensities
weighted by the solid angles
of annuli on the sky, is then
\begin{equation}
f(\lambda) = \int_0^\infty I(\lambda,y) \frac{2 \pi y}{ d^2 } dy
\end{equation}
where $d$ is the source distance.

\subsection{Continuum Lightcurves and Spectra}
\label{sec:thlght}

In a forthcoming paper (Pearson, Horne \& Skidmore, in prep.) we will 
consider generic analytic models for fireball behaviour applicable
to a variety of systems. We use the general opacity given in equation 
(\ref{eqn:genopac}).
We quote here an approximation to the lightcurve behaviour
being the sum of optically thick and optically thin contributions:
\begin{equation}
f_{\nu} = f_{\nu,0} \beta_{0}^{2} \tau_{0}^{b} s\left(\tau_{0}\right)
\end{equation}
where
\begin{eqnarray}
f_{\nu,0} & = & \frac{\pi a_{0}^{2}}{2 d^{2}} 
B_{\nu}\!(T), \\
\beta_{0} & = & \left[ \frac{\epsilon}{\nu^{3}} 
\frac{\kappa_{1} M^{2}}{\sqrt{2 \pi T_{0}} \pi^{3} a_{0}^{5}} 
\right]^{\frac{2}{10-\Gamma}}, \label{eqn:beta0} \\
\epsilon & = & 1-e^{-\frac{h \nu}{k T}} \\
\tau_0 & = & \left(\frac{\beta_{0}}{\beta}\right)^{\frac{10-\Gamma}{2}}, \\
b & = & \frac{6-\Gamma}{10-\Gamma}, \\
s(\tau) & = & \left\{ 
{\begin{array}{ll} 1 & \tau<1 \\ 
{\displaystyle \frac{1+\ln\tau}{\tau}} & \tau\geq1 \end{array}
} \right. .
\end{eqnarray}
Here $\beta_{0}$ is the time that the fireball first becomes optically thin 
through its centre and $\tau_{0}$ is the current optical depth through the 
fireball centre.

The peak flux occurs when
\begin{equation}
\tau_{0} = e^{\frac{b}{1-b}}
\end{equation}
which occurs at a time
\begin{equation}
\beta_{\rm pk} = \beta_0 e^{-\frac{b}{2}}.
\label{eqn:betapk}
\end{equation}
We note that equations~(\ref{eqn:beta0}) and (\ref{eqn:betapk}) predict 
that the
peak flux will occur at different times for different wavelengths.
In principle we could use this as a test for the effective $\Gamma$. However,
in practice, both the assumptions in our treatment and the necessary accuracy 
of observations, might make it difficult to detect the difference between 
$\beta_{\rm pk}\sim\lambda^{\frac{3}{5}}$ for an isothermal fireball and 
$\beta_{\rm pk}\sim\lambda^{\frac{3}{4}}$ for adiabatic cooling.
In contrast, the late time behaviour is predicted to have a more significant
dependance on $\Gamma$
\begin{equation}
f_{\nu} \sim \beta^{-\frac{6-\Gamma}{2}}.
\end{equation}
Coverage of a full flare may well make the $\beta\sim\beta^{-2}$ or 
$\beta^{-3}$ difference discernable.

\begin{figure}
\vspace{6.25cm}
\includegraphics{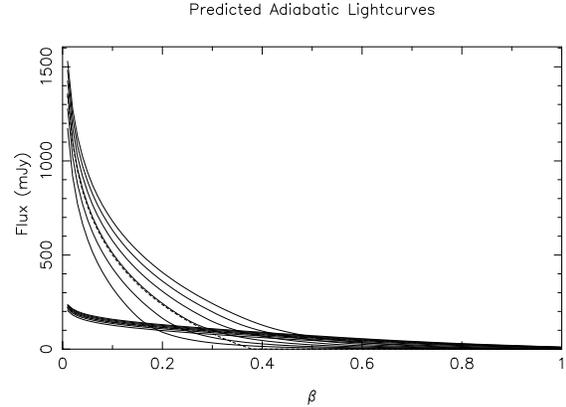}
\caption{The predicted fluxes at 
$3~000$ (lowest), $4~000$, $5~000$, $6~000$, $7~000$ {and} $8~000$~\AA~ 
(highest) for 
an adiabatic fireball with 
$M=10^{17}~{\rm kg}$ and $T_0=10~000~{\rm K}$. 
The upper family  
have $a_{0}=10^8~{\rm m}$ and the lower $a_{0}=3\times10^7~{\rm m}$. 
The dashed lines show the rapidly declining optically thick contribution
at $8~000$~\AA.}
\label{fig:adithecur}
\end{figure}

\begin{figure}
\vspace{6.25cm}
\includegraphics{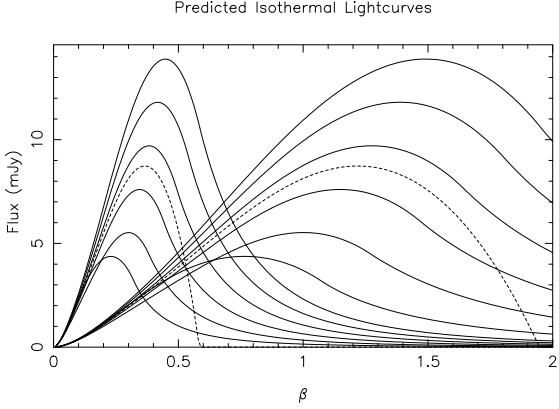}
\caption{The predicted fluxes at 
$3~000$ (lowest), $4~000$, $5~000$, $6~000$, $7~000$ {and} $8~000$~\AA~ 
(highest) for 
an isothermal fireball with 
$M=10^{17}~{\rm kg}$ and $T_0=10~000~{\rm K}$. The family on the 
left have $a_{0}=10^8~{\rm m}$ and the right $a_{0}=3\times10^7~{\rm m}$. 
The dashed lines show the optically thick contributions
at $8~000$~\AA}
\label{fig:isothecur}
\end{figure}

\begin{figure}
\vspace{6.25cm}
\includegraphics{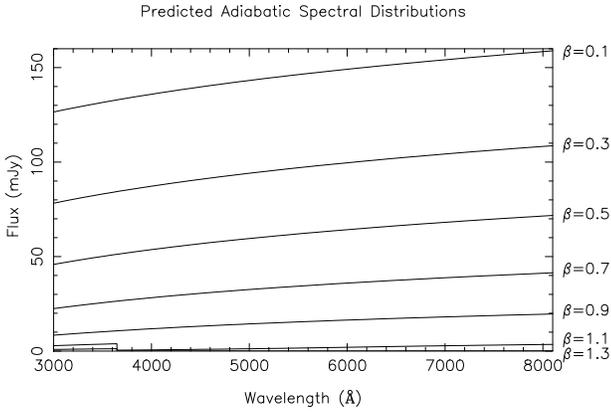}
\caption{The predicted spectral distribution for an adiabatic
fireball with $M=10^{17}~{\rm kg}$, $T_0=10~000~{\rm K}$ and 
$a_{0}=3\times10^7~{\rm m}$.}
\label{fig:adithespec}
\end{figure}

\begin{figure}
\vspace{6.25cm}
\includegraphics{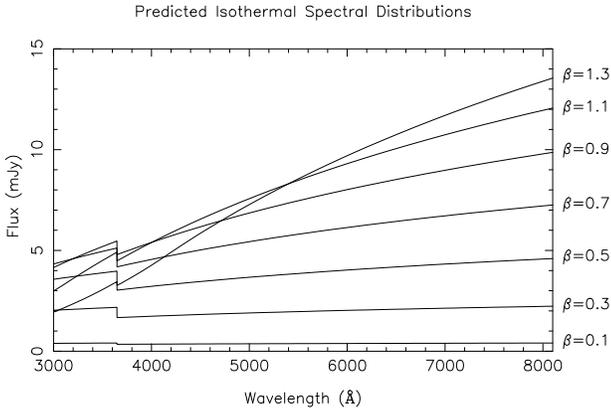}
\caption{The predicted spectral distribution for an isothermal fireball
with  $M=10^{17}~{\rm kg}$, $T_0=10~000~{\rm K}$ and 
$a_{0}=3\times10^7~{\rm m}$.}
\label{fig:isothespec}
\end{figure}

The total flux at several wavelengths and the optically thick contribution 
at $8~000$~\AA~are plotted in 
Figs.~\ref{fig:adithecur} and \ref{fig:isothecur} for 2
sets of physical conditions, using Population I composition and a distance of 
$100~\rm{pc}$. The corresponding spectral distributions for one of the 
parameter sets are plotted in
Figs.~\ref{fig:adithespec} and \ref{fig:isothespec}. We would expect the 
early and late time behaviours to be
accurately represented by this analysis, but the detailed behaviour when
both free-free and bound-free opacity sources are important is less exact.



The lightcurves for the isothermal models
show a rise to the peak value before turning over and undergoing a slower
decline. The predicted fluxes are very similar to observed fireballs and 
morphologically the similarity to observations is remarkable. It is also 
clear that the peak flux is independent of our choice of fiducial size.
We can see how 
the combination of the overall spectral slope and the large Balmer jump at 
early times leads to the $3~000$~\AA~curve exceeding that for $4~000$~\AA. As
ionization occurs due to the decreasing density, bound-free opacity 
becomes less important and the Balmer jump declines. The $3~000$~\AA~curve
then drops below the $4~000$~\AA~and behaves similarly to the other
curves in the family. The size of our Balmer jump, however, is still smaller 
than that observed suggesting that we may need to improve the treatment of 
opacity across this region.

The adiabatic lightcurves, however, show only a decline from
a large initial flux. We would expect the evolution
of the optically thick region here to be similar to that of the radiative
core photosphere ie. it would be initially advected outward with the flow 
before eventually turning over and collapsing back to zero size. The
emitted flux, however, does not follow this scheme as the fireball
is extremely hot at early times and the evolution of 
temperature and the blackbody intensity dominates the lightcurve evolution.
The spectra in this case only begin to show a Balmer jump late on in
the evolution once the fireball has cooled enough for significant recombination
to have taken place.

\subsection{Fireball Tomography}

\begin{figure*}
\begin{minipage}{17cm}
\vspace{10.5cm}
\includegraphics{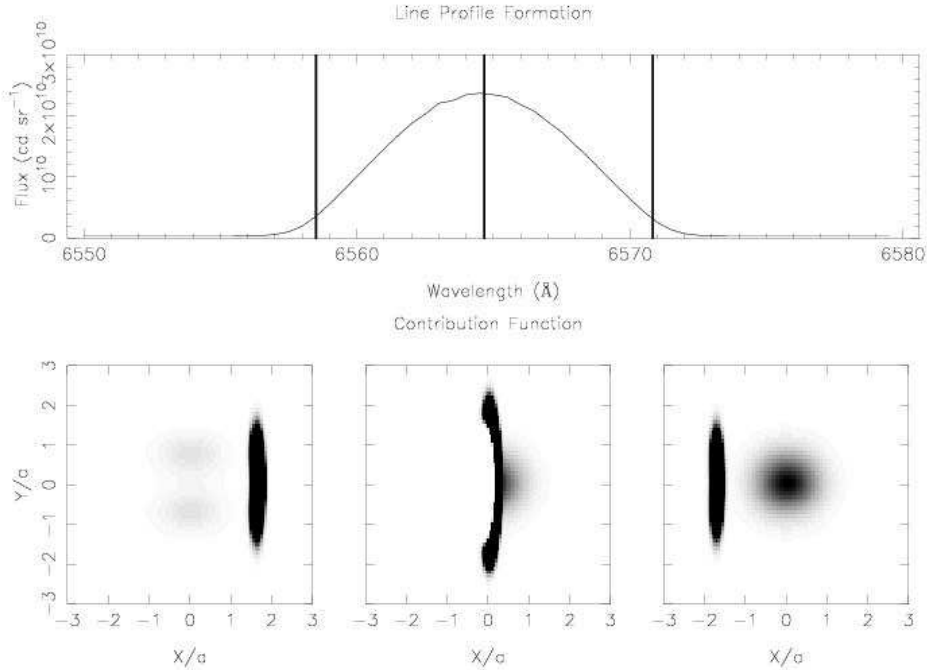}
\caption{Contribution to the flux at the wavelengths indicated from 
different positions in the fireball. The observer is at $+\infty$ on the $X$ 
axis and $Y$ is the impact parameter. Both axes are labelled in multiples of 
the current lengthscale $a=4.2\times10^{7}~{\rm m}$.}
\label{fig:isocont}
\end{minipage}
\end{figure*}

For our spherical expansion the component of velocity along the line of sight
is proportional to the distance from the centre towards the observer, ie.
$V_{X} = HX$. As a result, 
each velocity in the emission line profile originates from a unique 
surface perpendicular to the line of sight. 
This is illustrated in Fig.~\ref{fig:isocont},
which shows a side view of the fireball with the contribution to the flux at 
3 different velocities in the
H$\alpha$ profile. On the blue side of line centre, the resonant surface,
approximately perpendicular to the line of sight, is on the near side of the
fireball. The fireball is optically thick through its centre but sufficiently
transparent at larger impact parameters so that small continuum contributions
to the flux are made from  a peanut shaped region along the $y$-axis about 
the origin.
 
The resonant surface at the line centre passes through the centre of the 
fireball.
There is a continuum contribution from the rising density in front of the 
resonant plane around the origin and
then a line contribution from the resonant position. The larger optical depth
through the centre compared to larger impact parameters gives rise to the 
curved shape. The contribution from the back of the resonant region through
the line centre is unable to escape to the observer through the overlying 
material giving rise to the distorted shape. 

The flux on the red side of the line shows the two contributing regions
more clearly. A continuum contribution with the density peak and then another 
peak at the resonant position.

Such fireball tomography may allow us, in future, to probe the structure of 
a fireball using high-resolution spectra. For example, in 
Fig.~\ref{fig:isocont}, the Gaussian velocity profile arises from the 
assumed Gaussian density profile in equation~(\ref{denprof}). A different 
density profile would change the shape of the line profile.

\subsection{Numerical Considerations}

Our numerical integrations of equation (\ref{radtrans}) were carried out at 
appropriate
wavelength intervals to resolve the emission line profiles and with spatial
steps to reflect the variations in the density and ionization structures.
For conservative parameters $T=10^{4}~\mbox{K}$ and 
$\lambda_0=3~500~\mbox{\AA}$, the thermal line width is 
$\Delta\lambda\sim\frac{\lambda_0}{c}\left(\frac{k T}{m}\right)^{\frac{1}{2}}
 \approx 0.1~\mbox{\AA}$.
Opacity information was therefore generated and stored at wavelength and 
spatial intervals of $0.03~\mbox{\AA}$ and $0.1~a$ respectively. 
When generating the full spectra, intensities were calculated every 
$1~\mbox{\AA}$, giving a velocity resolution of 
$86~\mbox{km}~\mbox{s}^{-1}$ at $3~500~\mbox{\AA}$. 
The integration was carried out with spatial steps that were calculated to be 
either $\Delta \tau=0.005$ or $\Delta x =0.1~a$ whichever was shorter. 

In general the opacity at any position in the fireball consisted of
both continuum and line contributions. Continuum opacities were 
generally calculated using the methods of \scite{gray76} with the exception
of H$^{-}$ bound-free \cite{geltman62} and free-free \cite{stilley70}, 
He$^{-}$ \cite{mcdowell66} and HeI bound-free \cite{huang48}. 
Line opacities used oscillator strengths and energy level information
downloaded from the NIST atomic database\footnote{
http://physics.nist.gov/cgi-bin/AtData/main\_asd}.

We aim, in the longer term, to speed-up our code by using a Sobolev method, ie.
use the fact that velocity is a monotonically increasing function of position,
to find the resonant point in the fireball at each wavelength for a given 
line. We might also achieve time savings by identifying regions where no
lines make significant contributions to the flux. These methods would
enable us to use coarser grids or precalculated tables when appropriate. This
would simplify the integration scheme and possibly allow us to fit both the 
emission lines and continuum simultaneously with an amoeba code. 

\section{Comparison with Observations}
\label{sec:results}

\subsection{Observed Flare in AE Aqr}

\begin{figure}
\vspace{6.25cm}
\includegraphics{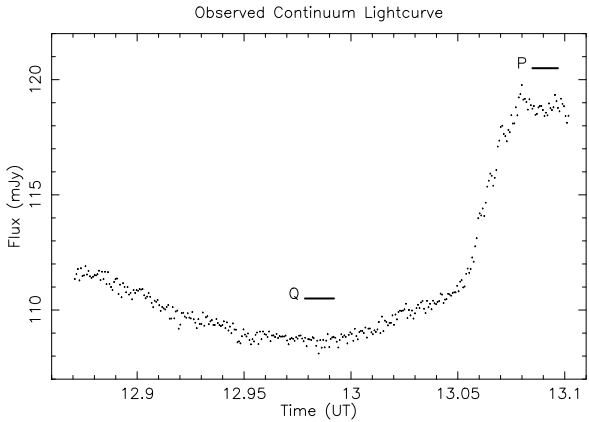}
\caption{Observed lightcurve reported in \protect\scite{skidmore02} showing
the decline of an old flare and start of a small new flare.}
\protect\label{fig:datlight}
\end{figure}

To test our fireball model, we aimed to reproduce the observations reported in 
\scite{skidmore02} of high time-resolution optical spectra of an AE Aqr flare 
taken with the Keck telescope. The lightcurve formed from these data is
plotted in Fig.~\ref{fig:datlight}. Although the dataset only lasts
around 13 minutes it shows the end of one flare and beginning of 
another. The major features of the flaring observed in the 
system are apparent in the lightcurve.

\begin{figure*}
\begin{minipage}{17cm}
\vspace{10.5cm}
\includegraphics{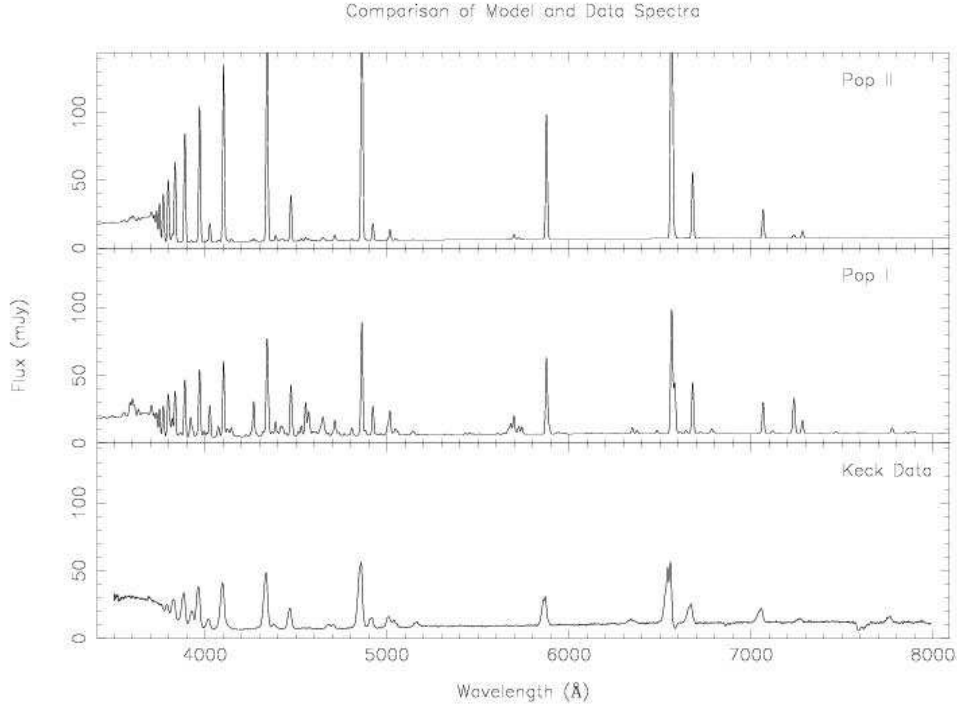}
\caption{Population I and II models compared to the observed data.}
\label{fig:popcomp}
\end{minipage}
\end{figure*}

We extracted the spectrum of the second flare by subtracting
the quiescent spectrum (Q) from the spectrum at the top of the new 
flare (P). The resultant optical spectrum is plotted in the bottom panel of
Fig.~\ref{fig:popcomp}.
Gaussian profiles fitted to several lines gave the parameters in 
Table~\ref{tab:expvel}.
Assuming the instrumental profile is also Gaussian, we remove the 
formal instrumental blurring ($9.8~\mbox{\AA}$ FWHM) to estimate 
the expansion velocity. 

These widths suggest a velocity dispersion of around $700~\mbox{km~s}^{-1}$.
Ascribing this to the velocity at 
$1\sigma$ of the density distribution gives a velocity of 
$\sim950~\mbox{km}~\mbox{s}^{-1}$ at $a_{0}$. Such a high velocity is 
larger than the expected $V\sim300~\mbox{km~s}^{-1}$ from Doppler tomography
\cite{welsh98} and may indicate that an additional, unknown source of 
blurring is present. Examination
of the arc line spectra support this suggestion. Since these were taken by 
flooding a large aperture with light it is not possible to accurately measure
the instrumental effect from these lines, but the line edges do suggest that 
the instrumental blurring is significantly larger than the formally derived
value.

\begin{table}
\begin{tabular}{rcc}
\hline
Line & Fitted FWHM & Deconvolved FWHM \\ 
     & (km $\mbox{s}^{-1}$) & (km $\mbox{s}^{-1}$) \\ \hline
H$\alpha$   & $1909\pm7$ & $1856\pm7$ \\
HeI 5876    & $1510\pm6$ & $1424\pm6$ \\
HeI 4922    & $1353\pm13$ & $1214\pm13$ \\
H$\beta$    & $1681\pm3$ & $1569\pm3$ \\
H$\gamma$   & $1743\pm3$ & $1589\pm3$ \\
H$\delta$   & $2171\pm4$ & $2050\pm4$ \\ 
H$\epsilon$ & $1645\pm5$ & $1469\pm5$ \\
CaH         & $2423\pm21$ & $2306\pm21$ \\ \hline
\end{tabular}
\protect\caption{Fitted and deconvolved widths of several emission lines.}
\protect\label{tab:expvel}
\end{table}

\subsection{Spectral Fit at Peak of Flare}

To estimate the fireball parameters $M$, $T$ and $a$ from the observed 
optical spectra we began with consideration of the continuum flux. We 
considered both Population I and Population II abundances. Population I 
abundances were set using the solar abundances presented by \scite{dappen00}.
For population II composition, we decreased from 0.085 to 0.028 
the ratio of the helium to hydrogen number densities and reduced the ratio 
for each metal abundance to 5\% of solar \cite{bowers84}. 
In either case we used an amoeba algorithm 
\cite{numrec} to optimize the fit by reducing $\chi^{2}$ to a minimum for the 
6 measured continuum fluxes in Table~\ref{contflux}. The best fit parameters
$M$, $T$ and $a$ and related secondary quantities are listed in 
Table~\ref{fitdata}. 

Using Population I composition, we derived 
parameters for the fireball at its peak of 
$M=3.7\times10^{16}$~kg, $a_0=5.1\times10^{7}$~m and $T_0=17~000$~K. 
Such a mass represents 
approximately 100~s of the L$_1$ mass transfer rate derived from 
equation (\ref{eqn:mdot}). Fig.~\ref{fig:popcomp} and Table~\ref{contflux}
show that these parameters provide a reasonably good fit to the observed 
Paschen continuum and Balmer Jump.

Population II composition produced best fit parameters of 
$M=6.8\times10^{16}$~kg, $a_0=9.6\times10^{7}$~m, $T_0=18~000$~K. 

Both compositions arrive at a model fit with remarkably similar calculated
fluxes which have a slightly steeper slope to the Paschen continuum than
the observations. The fits also produce a large Balmer jump but appears to
have difficulty in creating one quite as strong as that observed. 
A change in the 3rd significant figure of one of the fit parameters generally 
leads to a similar change in the calculated fluxes. Both sets of
parameters have similar temperatures but the Population II models require
roughly a factor 2 increase in mass and lengthscale equivalent to a factor
4 decrease in density. The central densities implied by both sets of
parameters are consistent with the typical values for the gas stream derived in
section~\ref{sec:estimate} and with the temperature range and emitting area
of \scite{beskrovnaya96}.

\begin{table*}
\begin{minipage}{150mm}
\begin{center}
\begin{tabular}{ccccc}
\hline
Range & Central Wavelength & Mean Flux & Pop I Fluxes & Pop II Fluxes \\ 
  (\AA)          & (\AA) & (mJy) & (mJy) & (mJy) \\ \hline
3602--3630 & 3616 & $30.68\pm0.20$  & 27.3 & 28.3\\
4150--4250 & 4200 & $6.812\pm0.019$ & 6.65 & 6.64\\
4520--4640 & 4580 & $7.442\pm0.017$ & 7.52 & 7.49\\
4735--4780 & 4758 & $7.394\pm0.028$ & 7.89 & 7.85\\ 
5400--5600 & 5500 & $9.114\pm0.016$ & 9.13 & 9.13\\
5600--5775 & 5688 & $9.346\pm0.018$ & 9.37 & 9.39\\ \hline
\end{tabular}
\end{center}
\caption{Mean flux measured in the 6 continuum regions with the
calculated continuum fluxes from the models.}
\protect\label{contflux}
\end{minipage}
\end{table*}

\begin{table}
\begin{center}
\begin{tabular}{rcc}
\hline
Quantity & Pop I & Pop II \\ \hline
$M~(10^{16}~{\rm kg})$ & $3.7$ & $6.8$ \\ 
$a_{0}~(10^{7}~{\rm m})$ & $5.1$ & $9.6$ \\
$T_{0}$~(K) & $17~000$  & $18~000$ \\
$v(a_{0})~({\rm km~s}^{-1})$ & $170$ & $170$\\
$1/H$~(s) & $300$ & $560$ \\
$\rho_{0}~(10^{-8}~{\rm kg~m}^{-3})$ & $5.1$ & $1.4$\\ \hline
\end{tabular}
\end{center}
\caption{Parameters used for the model spectra and useful secondary 
quantities.}
\protect\label{fitdata}
\end{table}

\begin{figure*}
\begin{minipage}{17cm}
\vspace{12cm}
\includegraphics{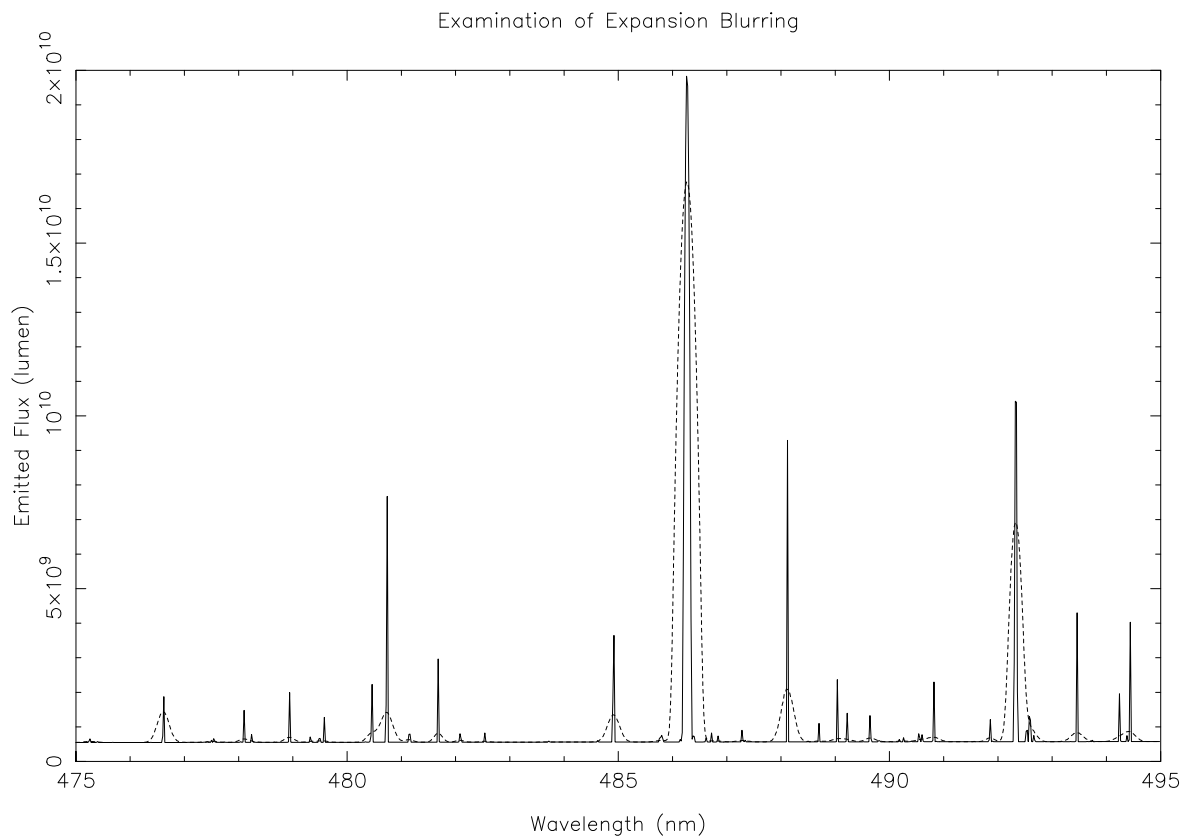}
\caption{Model spectra around H$\beta$ for Population I composition 
with zero expansion velocity (solid) and
an expansion velocity of $100~\mbox{km}~\mbox{s}^{-1}$ at $a_0$ (dotted)
before instrumental and distance effects are added.}
\label{fig:fullres}
\end{minipage}
\end{figure*}

Fig.~\ref{fig:fullres} shows the area around the H${\beta}$ line 
produced with the Population~I best fit parameters and with 
zero and modest expansion 
velocities, showing the intrinsic thermal resolution of the lines and the 
Doppler blurring due to the expansion before instrumental effects are added. 
Even with a relatively modest fiducial expansion velocity of 
$100~{\rm km~s}^{-1}$ we can see that many of the weaker lines are blurred into
obscurity simply by the Doppler effect. The  H${\beta}$ line at 
$4~861$~\AA~appears to be optically thick since its peak flux decreases only
slightly as the line width increases. In contrast, the ArII line at 
$4~807$~\AA~decreases greatly with Doppler broadening and forms a blend with 
the weaker blueward NII line indicating that it is optically thin.

\begin{figure}
\vspace{6.25cm}
\includegraphics{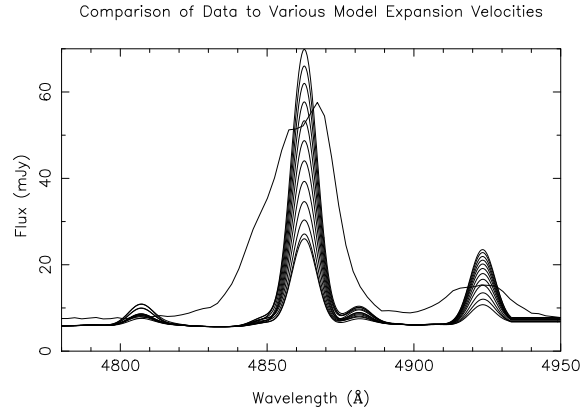}
\caption{Observed data around H$\beta$ plotted alongside model spectra with
Population I composition and expansion velocities at $a_0$ running from $0$ to 
$110~\mbox{km}~\mbox{s}^{-1}$ at $10~\mbox{km}~\mbox{s}^{-1}$ intervals.}
\label{fig:velcomp}
\end{figure}

In Fig.~\ref{fig:velcomp} we plot the observed data around H$\beta$ 
alongside Population I models generated with various expansion velocities. We  
consider the H$\beta$ region because the observed H$\alpha$ line
shows a complex structure. We blurred the model spectra 
with a Gaussian of $9.8~\mbox{\AA}$ FWHM and assumed a distance of $102$~pc 
\cite{friedjung97}. To match
the peak flux of H$\beta$ requires an expansion velocity of 
$75~\mbox{km}~\mbox{s}^{-1}$ at $a_0$. The HeI line at $4~923$~\AA~suggests an
expansion velocity of $40~\mbox{km}~\mbox{s}^{-1}$. The model lines are 
clearly narrower than the data. Considering this and the possibility of
additional, unidentified blurring suggests that a more reliable estimate 
might be the
integrated line flux. In this case, the H$\beta$ and HeI line suggest 
expansion velocities of $240~\mbox{km}~\mbox{s}^{-1}$ and
$100~\mbox{km}~\mbox{s}^{-1}$ respectively at $a_0$. 
Consequently, we adopt an expansion velocity at $a_0$ of 
$170~\mbox{km}~\mbox{s}^{-1}$ for the simulations.

The optical spectra produced with both compositions are compared to the 
observed mean spectrum at the flare peak in Fig.~\ref{fig:popcomp}. 
We see that the parameters, derived from the peak continuum fluxes alone, 
allow us to derive optical spectra with integrated line fluxes and ratios 
comparable to the 
observations. Both models have line widths narrower than the observations,
and hence peak line fluxes greater than those observed, which may result from
additional unidentified instrumental blurring. The Population I models show 
the flat Balmer decrement of saturated Balmer lines apparent
in the data. Comparing the lines that are present, however, particularly in
the $4000-5000$~\AA\ range, the Population II model appears to produce a much 
better fit: suppressing the metal
lines which do not appear in the observed spectra. Specifically, we note that 
SiIII$\lambda3080$, CII$\lambda4268$, 
the SiIII$\lambda\lambda4554,4555,4569,4576$ blend, 
OII$\lambda4650$, the SiI$\lambda5667$, SiIII$\lambda\lambda5697,5698,5741$ 
and ArII$\lambda5726$ blend and CII$\lambda\lambda7233,7238$ are present in 
the Population I models but are not present or are far
weaker in the data. Only the weak CII$\lambda3922$, CII$\lambda5147$, the 
SiII$\lambda\lambda6349,6373$ blend and the OI$\lambda7774,7775,7778$ blend 
are not reproduced by the Population II results.

\subsection{Lightcurves and Spectral Evolution}

The timescale for 
the evolution of the fireball implied by the observed lightcurve can be 
reproduced by the parameters derived solely from fitting to the 
peak spectrum. The evolutionary timescale $a_{0}/v(a_{0})=1/H$ is 
approximately $300$~s and $560$~s for Population I and II parameters
respectively. The encouraging result that these values are similar to the 
observed timescale suggests that our fireball models are on the right track.

\begin{figure*}
\begin{minipage}{17cm}
\vspace{10.5cm}
\includegraphics{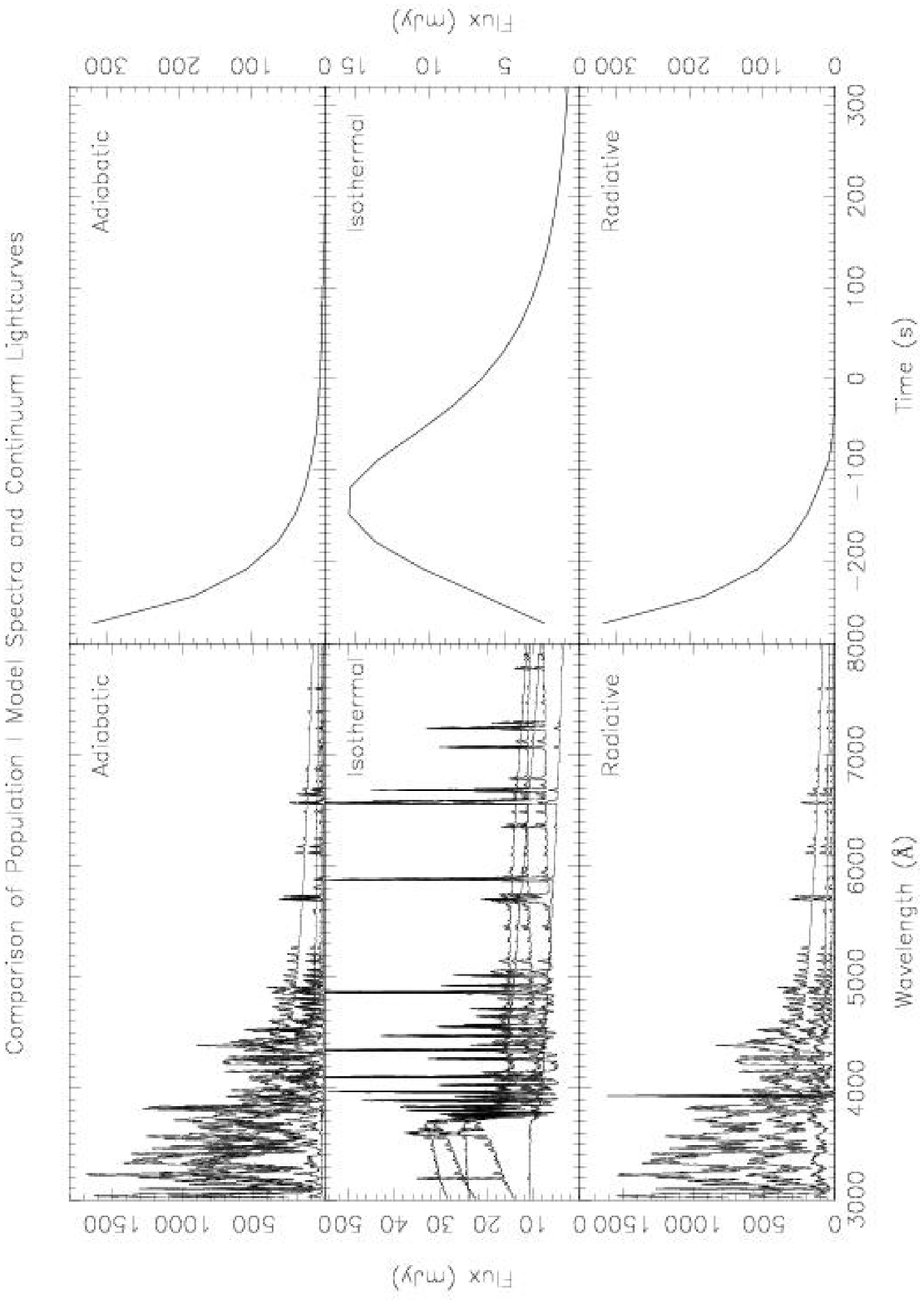}
\caption{Population I spectra at $t=-239,-179,-119,-60,0~\mbox{s}$, 
($\beta=0.2,0.4,0.6,0.8,1.0$) for the three cooling laws alongside their 
respective continuum lightcurves at $5~350$~\AA.}
\label{fig:popimods}
\end{minipage}
\end{figure*}

\begin{figure*}
\begin{minipage}{17cm}
\vspace{10.5cm}
\includegraphics{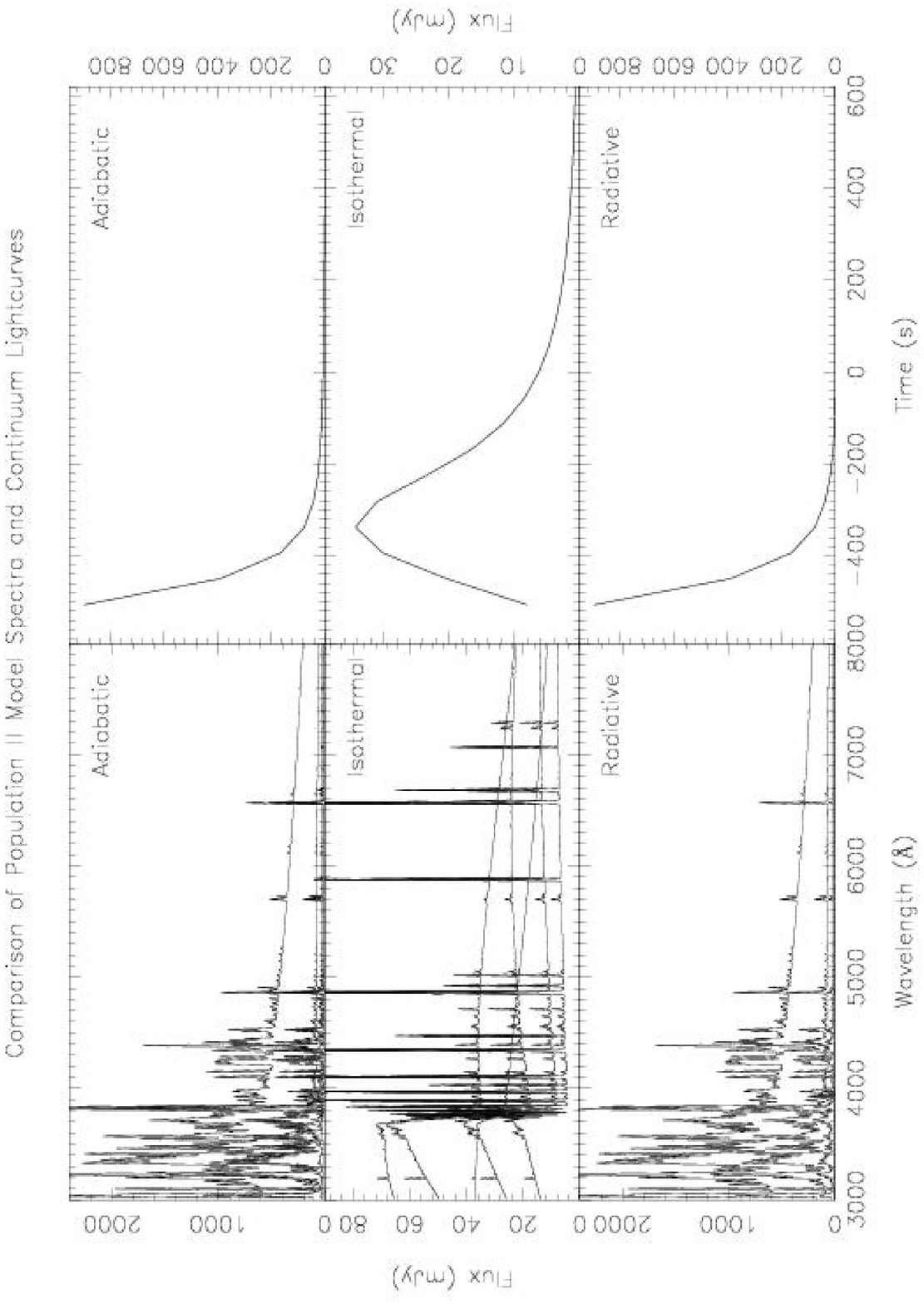}
\caption{Population II spectra at $t=-450,-337,-225,-113,0~\mbox{s}$, 
($\beta=0.2,0.4,0.6,0.8,1.0$) for the three cooling laws alongside their 
respective continuum lightcurves at $5~350$~\AA.}
\label{fig:popiimods}
\end{minipage}
\end{figure*}

We thus consider the time evolution of the fireball models in more detail.
Figs.~\ref{fig:popimods} and \ref{fig:popiimods}, for Population I and 
Population II abundances respectively, present the lightcurves and spectral 
evolution derived under the three assumptions we consider for the
thermal history of the fireball: adiabatic, isothermal, and radiative.  
In each case, the model fireball is constrained to evolve through the state
we found from the fit to the observations at the peak of the flare.
This moment is assigned to $t=0$ and hence our assignment of $0$ subscripts
for the fitted parameters above.


The isothermal models produces a lightcurve that rises and
falls, as observed.  Moreover, the peak in this lightcurve is
in fact reasonably close to the time when the spectrum matches the
spectrum observed at the peak of the flare.
The isothermal fireball spectra, both on the rise and on
the fall of the flare, exhibit strong Balmer emission
lines and a Balmer jump in emission.
Thus the isothermal fireball model provides a good fit not
only to the peak spectrum but also to the time evolution of
the AE Aqr flare.

The adiabatic and radiative models both fail to reproduce the
observed lightcurve or spectra.  The adiabatic model
matches the observed spectrum at one time,
but is too hot at early times and too cool at late times.
As a result the lightcurve declines monotonically from
early times rather than rising to a peak and then falling,
and the spectra have lines that do not match the observed spectra.
This disappointing performance is perhaps not too surprising,
since the temperature in this model falls through a wide range
while the observations suggest that the temperature is
always around $10^4$~K.

For the adiabatic fireball, equation~(\ref{initT}) implies that
the expansion rate is a measure of the initial temperature. Hence, for an 
adiabatic fireball we can derive its age from the time
taken to cool to the fitted temperature.
For the Population I parameters, $170~\mbox{km}~\mbox{s}^{-1}$ gives an 
initial temperature of $1.25\times10^{6}~\mbox{K}$, an age of $260~\mbox{s}$ 
and hence from (\ref{expfac}) an initial lengthscale of 
$5.9\times10^{6}~\mbox{m}$. The Population II
parameters give an age of $494~\mbox{s}$ and 
initial lengthscale of $1.1\times10^{7}~\mbox{m}$.
We note, however, that the lightcurve for the adiabatic fireballs 
do not show the rise to a peak for the same reason as the
theoretical curves in section~\ref{sec:thlght}. To 
explain the rise phase of the observed lightcurve we must interpret it as 
resulting from the collision of the initial gas blobs. Under this assumption, 
we can estimate the age and initial size {\it independently}, for the
adiabitic evolution,
directly from the data; knowing the rise time of the flare and the typical 
closing velocities:
\begin{eqnarray}
t_{\rm age} \sim & t_{\rm rise}  = & 63~{\rm s} \\
l \sim & V t_{\rm rise} = & 3\times10^{5} \times 60 = 1.8\times10^{7}~{\rm m}.
\end{eqnarray}

The radiative model looks roughly the same as the adiabatic
model.  This occurs because the cooling front (as shown
in Fig.~\ref{fig:tprof}) occurs well outside the photosphere.
We therefore ``see'' into the adiabatically-cooling core.
This is not consistent with our assumption that the
rapid cooling occurs near the photosphere as a result of
the opacity drop there and suggests that a more sophisticated approach
involving heating of the gas by emergent photons may be required for an
accurate treatment of such a model.   

\section{Discussion}
\label{sec:discuss}
The spectra produced with Population II composition reproduce the
optical observations well. The fitted parameters are consistent with both the 
expected conditions in the mass transfer stream and with the lower limits
on the density implied by the uv observations. The presence of high ionization
and semi-forbidden
lines in the uv data can be understood qualitatively in terms of their 
formation in the low
density outer fringes of the fireball and the wide variety of ionization 
states by the wide range of densities in the expanding fireball structure.
For $\beta=1$, $\mu=0.53$ and Population II parameters, the density
of $10^{17}~\mbox{m}^{-3}$, important for the uv semi-forbidden lines,
occurs at $\eta\approx2.3$. From Fig.~\ref{fig:ltelimtheory} we see that
this lies outside the limit for LTE behaviour.
Quantitative fits for the fringe and uv behaviour, therefore, remain to be 
addressed in the non-LTE regime extension to this work. 

Parameters implying a lower overall density
would result in an optically thinner fireball and spectrum. The Balmer 
decrement would increase as would the equivalent widths of the optically thick
Balmer lines as the continuum dropped away from the blackbody
envelope. Continued decrease would result in the lines eventually becoming 
optically thin and losing the characteristically consistent strengths present
when they are all saturated.

The behaviour of the lightcurve for an isothermal model, like its
theoretical counterpart is much
more similar to the observed lightcurve than the other models giving a peak 
close to the observed time without the need to invoke the collision process.
Clearly though, an expanding gas ball would be expected to cool both from
radiative and adiabatic expansion effects. 

The radiative model shows very similar behaviour to the adiabatic models. The
thin shell of cooling material at around $10^{4}~\mbox{K}$ does not provide 
sufficient opacity to 
mimic the isothermal behaviour and the adiabatic core is still visible
to the outside observer. This effect appears to be insensitive to the 
exact choice of temperature we assign to material emerging from the central
region.

We noted in section~\ref{sec:ionfreeze} that an accurate picture of the
ionization would need to follow the LTE to non-LTE ionization transition. 
However, even considering only the early evolution of the fireballs, the
isothermal lightcurve and spectral behaviour is stiller closer to the 
observations than the adiabatic.

In short, we find that isothermal fireballs reproduce the observations rather 
well, 
whereas adiabatically cooling fireballs fail miserably! How can the expanding 
gas ball present a nearly constant temperature at its surface?

\subsection{Thermostat}

We offer two possible mechanisms which may operate to maintain the apparent 
fireball temperature in the $1$--$2\times10^4~\mbox{K}$ region. The first 
relies on the fact that both free-free and bound-free opacity decrease with 
temperature and thus opacity peaks just above the temperature at which 
hydrogen begins to recombine.
In the model where a core region is cooling through radiation from its 
surface, we can envision a situation where the material just outside the core 
has reached this high opacity regime and is absorbing significant amounts of 
energy from the core's radiation field. This would help to counteract the 
effect of adiabatic cooling and may maintain an effectively isothermal blanket 
around the core. If the temperature in the blanket were to rise, the opacity 
would decrease, more energy would escape and the
blanket would cool again. Similarly, until recombination, cooling would
result in higher opacity and therefore greater heating of the blanket region.
For such a thermostatic mechanism to work, the optical depth through the
 blanket region would need to be high which would be consistent with it being 
the photosphere as seen by an outside observer. Realistic simulations of such a
model would require a detailed treatment of the radiation field and its 
heating effect using a more sophisticated method than the one we have 
employed here.

\subsection{External Photoionization}

\begin{figure}
\vspace{6.25cm}
\includegraphics{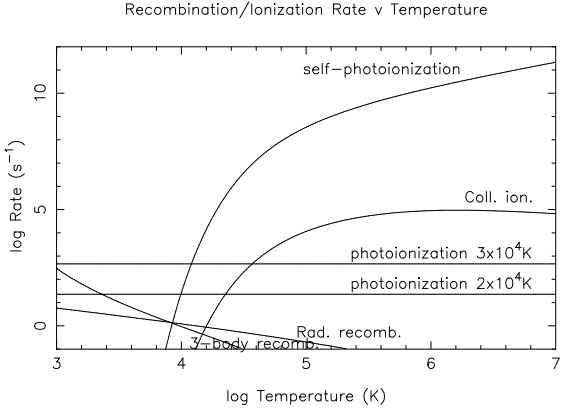}
\caption{Ionization and recombination rates for a hydrogen atom or ion 
respectively with $n_{\rm e}=3\times10^{18}~\mbox{m}^{-3}$.}
\label{fig:ionrate}
\end{figure}

Alternatively the ionization of the fireball might be held at a temperature
typical of the  $1$--$2\times10^4~\mbox{K}$ range through photoionization by 
the  white dwarf. In
Fig.~\ref{fig:ionrate} we plot the hydrogen ionization and recombination 
rates as a function of temperature for a typical electron number density 
$n_{\rm e}=3\times10^{18}~{\rm m}^{-3}$ 
using additional routines for photoionization 
from Verner et al. (1996) and collisional ionization by Verner using the 
results of Voronov (1997). We calculate a conservative overestimate for
the self-photoionization of the fireball assuming the sky is half filled by
blackbody radiation at the given fireball temperature. Also shown is the
photoionization due to a blackbody with the same radius and distance 
as the white dwarf. The general white dwarf temperature is uncertain but,
as mentioned earlier, there is evidence to suggest that the 
hot spots have a temperature of around $3\times10^{4}~\mbox{K}$. We can see 
that the flux from the white dwarf overtakes that from the fireball itself as 
the dominant ionization mechanism at $\sim10^{4}~\mbox{K}$. We have assumed
a typical distance from the white dwarf for the fireball equal to that of 
the L1 point from the white dwarf (although in a different direction). 
The recombination 
rates for this plot have been calculated assuming $n_{\rm e}$ is a constant 
and so for temperatures below about
$10^{4}~\mbox{K}$ are conservative overestimates of the LTE recombination 
rate that will occur in our fireball. In spite of this,
the white dwarf photoionization at $3\times10^{4}~\mbox{K}$ 
comfortably exceeds the 
recombination rates down to at least $10^{3}~\mbox{K}$ and, hence, we would 
expect the fireball to remain almost completely ionized in this case.
This suggests that the white dwarf radiation field may be the cause of the 
fireball appearing isothermal when we consider the
lightcurve behaviour and consistent line strengths and ratios. 

Ionization by an external black body at a different temperature will
clearly cause our fireball to no longer be in LTE and require a
non-LTE model to follow the fireball behaviour. However, we can anticipate
the results of a full non-LTE treatment as follows.
Writing the {\it total} ionization rate per atom per second from all processes 
as 
$b(n_{\rm H},n_{\rm e},T)$ and similarly the {\it total} recombination rate as 
$a(n_{\rm e},T)$, we have the simple equilibrium condition
\begin{equation}
b n_{\rm n} = a n_{\rm i}.
\end{equation}
Hence the ionization fraction is given by
\begin{equation}
\frac{n_{\rm i}}{n_{\rm n}+n_{\rm i}} = \frac{b}{b+a}.
\end{equation}

\begin{figure}
\vspace{6.25cm}
\includegraphics{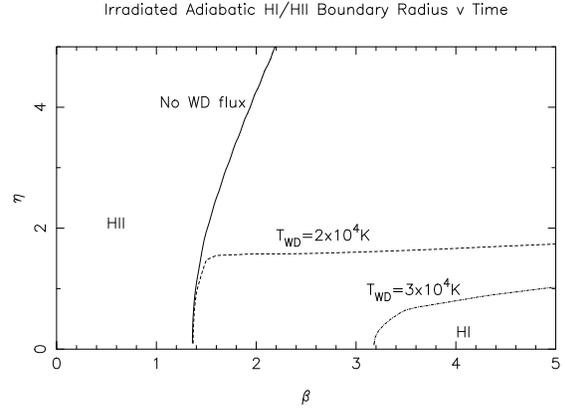}
\caption{Comparison of the evolution of the hydrogen ionization boundary
for purely LTE and LTE plus white dwarf photoionization cases.}
\label{fig:bndcmp}
\end{figure}

We calculate the evolutions of the boundary between HI and HII dominated 
regions for a pure hydrogen fireball using $M=10^{16}~\mbox{kg}$,
$a_{0}=10^{7}~\mbox{m}$ and $T_{0}=18~000~\mbox{K}$. 
Since in this simplified case $n_{\rm e}=n_{\rm i}$, we can easily iterate to 
produce self-consistent ionization profiles in cases where b contains 
contributions from ordinary atomic processes and also an additional 
contribution from white dwarf photoionization. We plot the evolutions in 
Fig.~\ref{fig:bndcmp} for no white dwarf photoionization and with rates
appropriate to photoionization from a white dwarf at 
$2\times10^{4}~\mbox{K}$ and $3\times10^{4}~\mbox{K}$. We see that the white 
dwarf can have a significant effect on the ionization structure of the 
fireball and keep large parts of it in an ionized state. 


\section{Summary}
\label{sec:summary}

We have shown for AE~Aqr how the observed flare spectrum and evolution is 
reproducible with an isothermal fireball at $T=18~000$~K with Population II 
abundances but not when adiabatic cooling is incorporated. We suspect that 
the cause of the apparently isothermal nature is a combination of two 
mechanisms. First, a nearly isothermal
photosphere which is self-regulated by the temperature dependance of 
the continuum opacity and the hydrogen recombination front and second, 
particularly at late times, by the ionizing effect of the white dwarf
radiation field. 

The observed photosphere in the isothermal models is 
initially advected outward with the flow before
the decreasing density causes the opacity to drop and the photosphere to 
shrink to zero size. This gives rise to a lightcurve that rises and then falls.
Emission lines and edges arise because the photospheric radius is larger at 
wavelengths with higher opacity. Surfaces of constant Doppler shift are
perpendicular to the line of sight and thus Gaussian density profiles give rise
to Gaussian velocity profiles.

Direct application of our LTE fitting method to the uv region would not 
reproduce the observed spectra with their variety of ionization states and 
semi-forbidden and permitted lines. We have
outlined techniques and possible improvements to the model that will increase 
our understanding further and enable us to take the work into the uv. More 
detailed modelling of the 
ionization and recombination processes is suggested that also incorporates the 
heating effects of both the white dwarf's and the fireballs own photon fields. 
The self-consistent solution to these rate equations will
extend the treatment to the non-LTE regimes discussed in 
Section~\ref{sec:discuss} and Appendix~\ref{sec:ltevalidity}. These 
modifications may in turn have consequences for the predicted optical spectra.

Improved modelling of these fireballs offers us the chance to probe the
chemical composition of the secondary star in AE Aqr. Normally this is 
difficult because the spectrum of the relatively dim secondary star 
is contaminated by light
from other components in the system. We expect that the fireball models will
be applicable more generally to the flickering observed in most CV
systems. 

\subsection*{Acknowledgements}

We would like to thank Gary Ferland and Kirk Korista for informative
discussions regarding atomic ionization and recombination processes and the 
routines for calculating them. We are grateful to the anonymous referee
and editor for helpful comments that improved the presentation
of this paper.

\appendix

\section{Validity of the LTE Approximation}
\label{sec:ltevalidity}

LTE requires that the electron, ions and photons are all in thermal 
equilibrium. LTE also requires that detailed balance is able to be 
maintained. Rapid expansion can interfere with this. Since recombination rates
decrease with
density, ionization equilibrium should break down in the low density 
outskirts of the fireball. 

We derive a limiting radius inside which statistical equilibrium for ionization
holds from the condition that the
expansion timescale ($t_{\rm exp}$) be longer than the recombination timescale
($t_{\rm rec}$). Now,
\begin{equation}
t_{\rm exp}=\frac{V}{\dot{V}}=\frac{\frac{4}{3}\pi r^{3}}{4\pi r^{2}\dot{r}}
           =\frac{r}{3\dot{r}} = \frac{\beta}{3H}
\label{eqn:exptscale}
\end{equation}
and
\begin{equation}
t_{{\rm rec},i} = \left\lbrace \alpha_{i} n_{\rm e}
\left[\frac{n_{i+1}}{n_{i}} - \frac{\alpha_{i-1}}{\alpha_{i}} 
\right]\right\rbrace^{-1} ,
\label{trecomb}
\end{equation}
where we have taken the effective recombination timescale for the $i$th state 
of a given element from Bottorff et al. (2000), 
$\alpha_{i}$ is the recombination coefficient
to the $i$th ionization state and the term in square brackets contains a 
correction factor to the commonly used expression. Here, we assume
that this term is unity but for detailed checks in section~\ref{detcheck}
we evaluate using the calculated densities.

\begin{figure}
\vspace{6.25cm}
\includegraphics{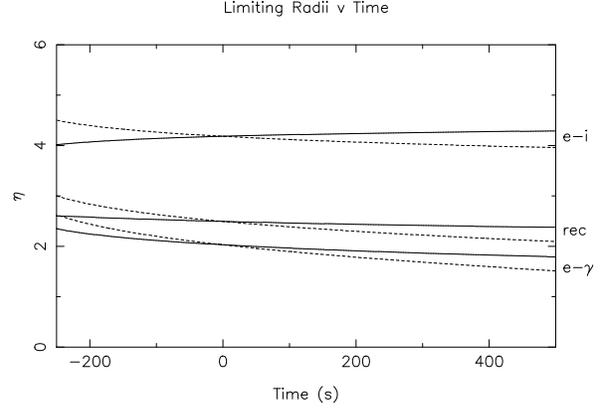}
\caption{Equilibrium limiting radii as a function of time for a pure
hydrogen fireball. Adiabatic evolutions are shown in solid lines, isothermal 
in dashed.}
\label{fig:ltelimtheory}
\end{figure}

For an initial estimate, we consider a fireball which is composed purely of
hydrogen.
Equating (\ref{eqn:exptscale}) and (\ref{trecomb}),
we plot an estimate of the limiting ionization equilibrium radius for hydrogen
in Fig.~\ref{fig:ltelimtheory}, using parameters 
$M=10^{17}~\mbox{kg}$, $a_{0}=10^{8}~\mbox{m}$,
$T_{0}=18~000~\mbox{K}$ and a velocity at $a_0$ of $300~{\rm km~s}^{-1}$
giving $H=3.0\times10^{-3}~\mbox{s}^{-1}$. 
We use a radiative recombination rate given by 
\begin{equation}
\alpha= a_{\rm r} \left[
\sqrt{\frac{T}{T_{1}}}\left(1+\sqrt{\frac{T}{T_{1}}}\right)^{1-b}
\left(1+\sqrt{\frac{T}{T_{2}}}\right)^{1+b}
\right]^{-1}
\label{eqn:radrecrate}
\end{equation}
where $a_{\rm r}=7.982\times10^{-17}~\mbox{m}^{3}~\mbox{s}^{-1}$, $b=0.7480$,
$T_{1}=3.148~\mbox{K}$ and $T_{2}=7.036\times10^{5}~\mbox{K}$
\cite{verner96}.
We can estimate the range of times for which the fireball is optically thick
using a general opacity
\begin{equation}
\kappa=\kappa_1 \epsilon T^{-\frac{1}{2}} \nu^{-3} \rho^{2} 
\label{eqn:genopac}
\end{equation}
where $\epsilon$ is the stimulated emission correction
and $\kappa_{1}$ is a constant dependant upon the composition
and whether bound-free or free-free opacity is being considered. For
free-free opacity and pure hydrogen composition 
$\kappa_{1}=1.34\times10^{52}$. We parameterize
the temperature evolution in the form $T=T_{0}\beta^{-\Gamma}$.
Thus, $\Gamma=0$ for an isothermal model and $\Gamma=2$ for adiabatic
cooling. Integrating (\ref{eqn:genopac}) over a line of sight through the 
centre of the
Gaussian density profile we can derive an optical depth through the fireball
\begin{equation}
\tau=\frac{\kappa_{1} \epsilon M^{2}}
{T_{0}^{\frac{1}{2}} \nu^{3} \pi^{\frac{5}{2}} a_{0}^{5}} 
\beta^{-\frac{10-\Gamma}{2}}. \label{eqn:genopdpth}
\end{equation}
Hence the condition for $\tau>1$ becomes
\begin {equation}
\beta \lta \left\{ 
{\begin{array}{ll} 0.35 & \mbox{for}~\Gamma=0 \\ 
0.27 & \mbox{for}~\Gamma=2 \end{array}
} \right.
\end{equation}
using the parameters considered above and evaluating at $3~000$~\AA.
We can see that the expanding gas is optically thick at early times and that
the limiting $\beta$ is relatively insensitive to the exact parameters given 
the strong power of $\beta$ in (\ref{eqn:genopdpth}). However, it is most 
strongly dependant on (at least inversely proportional to) the fiducial
lengthscale $a_{0}$.

We can estimate the time for temperature equilibrium to be established
from the collision frequency between electrons and ions using the method of
\scite{goldston95}
\begin{eqnarray}
\nu_{\rm ei} & = & n_{\rm i} <\!\sigma_{\rm ei} {v_{\rm e}}\!> \\
             & = & \frac{2^{\frac{1}{2}} n_{\rm i} Z^{2} e^{4} \ln \Lambda}
{12 \pi^{\frac{3}{2}} \epsilon_{0}^{2} m_{\rm e}^{\frac{1}{2}} 
(kT_{\rm e})^{\frac{3}{2}}} \\
& \approx & 7.3\times10^{-5}~\frac{n}{T_{\rm e}^{\frac{3}{2}}}~\mbox{s}^{-1}\\
& \approx & 3.3\times10^{8} \beta^{3(\frac{\Gamma}{2}-1)} 
e^{-\eta^{2}}~\mbox{s}^{-1}.
\end{eqnarray}
where we have evaluated the expression for the case of hydrogen using 
$\ln\lambda=20$ and the same fiducial parameters as above.
By considering the energy exchanged in a collision and integrating over a
Maxwellian distribution of electron velocities we arrive at a timescale
for ions and electrons to come into equilibrium
\begin{eqnarray}
t_{\rm ei,eq} & = & \frac{3 \pi (2 \pi)^\frac{1}{2} \epsilon_{0}^{2} m_{\rm i} 
(kT_{\rm e})^{\frac{3}{2}}}
{n_{\rm e} Z^{2} e^{4} m_{\rm e}^{\frac{1}{2}} \ln \Lambda} \\
& \approx & 2.8\times10^{-6} \beta^{3(1-\frac{\Gamma}{2})} 
e^{\eta^{2}}~\mbox{s}.
\end{eqnarray}
Electrons collide on a similar frequency but share out there energy on a 
timescale roughly $\left(\frac{m_{\rm e}}{m_{\rm H}}\right)^{\frac{1}{2}}$ 
quicker.

We can estimate the timescale for photons and electrons to reach
equilibrium in an analogous way using the free-free cross-section 
$\sigma_{\rm ff}$ and the opacity outlined above
\begin{eqnarray}
\nu_{\gamma e} & = & n_{e} \sigma_{\rm ff} c \\
               & = & \kappa_{1} \epsilon \frac{ \rho^{2} c}
{\nu^{3} T_{\rm e}^{\frac{1}{2}}}  \\
               & \approx & 2.9\times10^{-3} \beta^{\frac{\Gamma}{2}-6} 
e^{-2\eta^{2}}~\mbox{s}^{-1} \label{eqn:nueg}
\end{eqnarray}
where we have evaluated the expression in the uv at $2~000$~\AA. 
This estimate becomes less accurate at low optical depths since photons will
be able to travel through significant changes in density before absorption.
This would necessitate an integral over the probability of absorption per unit
distance for an exact treatment. However, since inward photons will approach 
regions of higher density and outward lower density the mean density 
represented by this method is sufficiently precise for the use here. Once 
the expansion becomes optically thin, then the assumption of a Planck 
distribution in the derivation of (\ref{eqn:nueg}) breaks down and the 
definition of the limiting radius itself becomes more fuzzy as photons will 
interact with electrons throughout the volume. When a 
photon is absorbed it gives up all its energy so that we can calculate the 
timescale to approach equilibrium by a similar method to that above. 
Integrating over frequency we arrive at a timescale 
\begin{eqnarray}
t_{\gamma,{\rm e}} & = & \frac{3 c^{2} T_{{\rm e}}^{\frac{1}{2}}}
{16 \pi m_{\rm H}^{2} \kappa_{1} n_{\rm i}}\\
 & \approx & 1.8 \beta^{3-\frac{\Gamma}{2}} e^{\eta^{2}}~\mbox{s}. 
\end{eqnarray}

The limiting radii, where these timescales equal the expansion timescale are
plotted as a function of time in Fig.~\ref{fig:ltelimtheory} alongside
the ionization equilibrium limiting radius. The lowest of these three 
boundaries marks the limit for LTE to hold.

Even for the late time behaviour, the minimum photon-electron 
equilibrium radius occurs at $R\sim1.5a_{0}$ ie. $2.0\sigma$
which contains $\sim95$\% of the mass. The ionization-equilibrium timescale
is the most volatile of the plotted timescales, since the limiting radius 
depends on 
the ionization structure for each element. As such, it should be calculated 
after each 
simulation using the stored ionization history to double check the validity. 
This radius will also evolve differently for each element.

\subsection{Detailed Check}
\label{detcheck}

\begin{figure}
\vspace{6.25cm}
\includegraphics{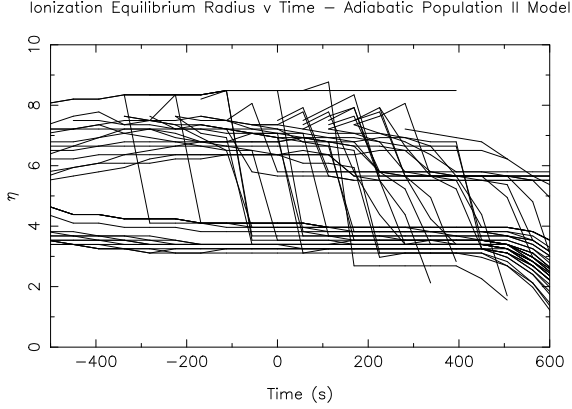}
\caption{Ionization equilibrium limit of the simulations for a 
Population II, adiabatically
cooling fireball for all ions included with non-zero densities.}
\label{fig:adiltelim}
\end{figure}

\begin{figure}
\vspace{6.25cm}
\includegraphics{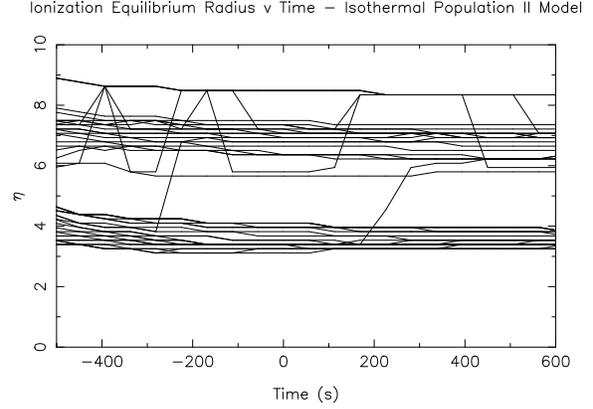}
\caption{Ionization equilibrium limit of the simulations for a 
Population II, isothermal
fireball for all ions included with non-zero densities.}
\label{fig:isoltelim}
\end{figure}

Examining the conditions for equality of equations 
(\ref{eqn:exptscale}) and (\ref{trecomb}), we can check the limiting
radius for ionization equilibrium to hold for each ion included in each of the 
simulations.
We plot results, for all the included ions, for the Population II adiabatic 
and isothermal models in Figs.~\ref{fig:adiltelim} and \ref{fig:isoltelim}.
We use routines to calculate the recombination coefficient for all the 
appropriate processes using codes for radiative events by Verner 
\cite{verner96,pequignot91,arnaud92,shull82,landini90,landini91}, 
dielectronic \cite{mazzotta98},
three-body by Cota (1987) as used in Cloudy and charge transfer 
recombination events by Kingdon \& Ferland (1996). 
We can see that, in general, there is a slow evolution of the limiting radius
with time. The quantised nature of the grid on which the ionization states,
and thus recombination timescales, were calculated is apparent 
($\Delta\eta=0.1$). The dramatic switching that sometimes occurs results from 
changes in the ionization structure between one spectrum and the next. However,
since we only require enough spectra to generate a clear lightcurve, the lack 
of temporal resolution for these changes is not of significant concern.

Except for late times, both of these evolutions give an ionization 
equilibrium limit 
$\eta\ga3.0$ ie. $4.2\sigma$ of the density distribution. As a result
we can be assured that, under our assumptions, only a small fraction of the 
mass of the fireball
has non-LTE composition and our LTE treatment is reasonable.


\bsp


\begin{thebibliography}{}
  \bibitem[\protect\citename{Abada-Simon et al.} 1995]{abada-simon95}
    Abada-Simon, M., Bastian, T.\ S., Horne, K.\ D., Robinson, E.\ L., 
    Bookbinder, J.\ A., 1995, in Buckley, D.\ A.\ H., Warner, B., eds.,
    Proc. Cape Workshop on Magnetic Cataclysmic Variables, ASP Conf. Series,
    355 
  \bibitem[\protect\citename{Arnaud \& Raymond} 1992]{arnaud92}
    Arnaud, M., Raymond, J.,
    1992, ApJ, 398, 394 
  \bibitem[\protect\citename{Bak} 1996]{bak96}
    Bak, P.,
    1996, How nature works: the science of self-organized criticality, 
    Copernicus, New York
  \bibitem[\protect\citename{Bastian, Dulk \& Chanmugam} 1988a]{bastian88a}
    Bastian, T.\ S., Dulk, G.\ A., Chanmugam, G.
    1988, ApJ, 324, 431 
  \bibitem[\protect\citename{Bastian, Dulk \& Chanmugam} 1988b]{bastian88b}
    Bastian, T.\ S., Dulk, G.\ A., Chanmugam, G.
    1988, ApJ, 330, 518 
  \bibitem[\protect\citename{Beskrovnaya {\rm et~al.}} 1996]{beskrovnaya96}
    Beskrovnaya~N., Ikhsanov~N., Bruch~A., Shakhovskoy~N., 
    1996, A\&A, 307, 840
  \bibitem[\protect\citename{Bowers \& Deeming} 1984]{bowers84}
    Bowers, R.\ L., Deeming, T., 1984, Astrophysics I Stars, 
    Jones and Bartlett, Boston 
  \bibitem[\protect\citename{Bruch }1991]{bruch91}Bruch~A., 1991, A\&A, 251,
    59
  \bibitem[\protect\citename{Bruch }1992]{bruch92}
    Bruch, A., 1992, A\&A, 266, 237 
  \bibitem[\protect\citename{Bruch \& Grutter }1997]{bruch97}Bruch~A.,
    Gruuter~M., 1997, Act Ast, 47, 307
  \bibitem[\protect\citename{Cota} 1987]{cota87}
    Cota, S.\ A., 1987, Ph.D. Thesis, Ohio State Univ.
  \bibitem[\protect\citename{Dalgarno \& McCray} 1972]{dalgarno72}
   Dalgarno, A., McCray, R.\ A., 1972,
   ARA\&A, 10, 375  
  \bibitem[\protect\citename{D\"{a}ppen }2000]{dappen00}
    D\"{a}ppen, W.,
    2000, in Cox, A.\ N., Allen's Astrophysical Quantities, Chapter 3, 
    Springer-Verlag, New York
  \bibitem[\protect\citename{de Jager} 1994]{dejager94}
   de Jager, O.\ C., Meintjes, P.\ J., O'Donoghue, D., Robinson, E.\ L., 1994,
   MNRAS, 267, 577  
  \bibitem[\protect\citename{Elsworth \& James }1982]{elsworth82}
   Elsworth, Y.\ P., James, J.\ F., 1982, MNRAS, 198, 889  
  \bibitem[\protect\citename{Eracleous et al. }1994]{eracleous94}
   Eracleous, M., Horne, K.\ D., Robinson, E.\ L., Zhang, E.-H., Marsh, T.\ R.,
   Wood, J.\ H., 1994, ApJ, 433, 313 
  \bibitem[\protect\citename{Eracleous \& Horne }1996]{eracleous96}
   Eracleous, M., Horne, K.\ D., 1996, ApJ, 471, 427  
  \bibitem[\protect\citename{Frank, King \& Raine } 1992]{FKR}
    Frank, J., King, A.\ R., Raine, D.\ J.,
    1992, Accretion Power in Astrophysics, Cambridge Univ. Press
  \bibitem[\protect\citename{Friedjung} 1997]{friedjung97}
    Friedjung, M., 1997, NewA, 2, 319
  \bibitem[\protect\citename{Geltman} 1962]{geltman62}
    Geltman, S., 1962, ApJ, 136, 935 
  \bibitem[\protect\citename{Glasco \& Zirin }1964]{glasco64}
    Glasco, H.\ P., Zirin, H., 1964, ApJS, 9, 193
  \bibitem[\protect\citename{Goldston \& Rutherford }1995]{goldston95}
    Goldston, R.\ J., Rutherford, P.\ H.,
    1995, Introduction to Plasma Physics, Bristol, IOP Publishing
  \bibitem[\protect\citename{Gray }1976]{gray76}
    Gray, D.\ F.,
    1976, Observations and Analysis of Stellar Photospheres, New York, Wiley
  \bibitem[\protect\citename{Horne }1999]{horne99}
    Horne, K.\ D.,
    1999, in Hellier, K., Mukai, K., eds, 
    Annapolis Workshop on Magnetic Cataclymic Variables, ASP Conf. Series,
    157, 357
  \bibitem[\protect\citename{Huang} 1948]{huang48}
    Huang, S., 1948, ApJ, 108, 354
  \bibitem[\protect\citename{Hameury, King \& Lasota} 1986]
   {hameury86} 
   Hameury, J.\ M., King, A.\ R., Lasota, J.-P., 1986,
   A\&A, 162, 71
  \bibitem[\protect\citename{Jameson, King \& Sherrington }1980]{jameson80} 
   Jameson, R.\ F., King, A.\ R., Sherrington, M.\ R., 1980,
   MNRAS, 191, 559  
  \bibitem[\protect\citename{Kingdon \& Ferland} 1996]{kingdon96} 
   Kingdon, J.\ B., Ferland, G.\ H., 1996,
   ApJS, 106, 205  
  \bibitem[\protect\citename{Kovetz, Prialnik \& Shara} 1988]{kovetz88} 
   Kovetz, A., Prialnik, D., Shara, M.\ M., 1988, ApJ, 325, 828
  \bibitem[\protect\citename{Kuijpers et al.} 1997]{kuijpers97} 
   Kuijpers, J., Fletcher, L., Abada-Simon, M., Horne, K.\ D.,
   Raadu, M. A., Ramsay, G., Steeghs, D., 1997, A\&A, 322, 242
  \bibitem[\protect\citename{Landini \& Monsignori Fossi} 1990]{landini90}
    Landini, M., Monsignori Fossi, B.\ C., 1990,
    A\&AS, 82, 229
  \bibitem[\protect\citename{Landini \& Monsignori Fossi} 1991]{landini91}
    Landini, M., Monsignori Fossi, B.\ C., 1991,
    A\&AS, 91, 183
  \bibitem[\protect\citename{Lynden-Bell \& Tout} 2001]{lyndenbell01}
    Lynden-Bell, D., Tout, C., 2001,
    ApJ, 558, 1
  \bibitem[\protect\citename{McDowell, Williamson \& Myerscough} 1966]
   {mcdowell66} 
   McDowell, M.\ R.\ C., Williamson, J.\ H., Myerscough, V.\ P., 1966,
   ApJ, 144, 827  
  \bibitem[\protect\citename{Mazzotta et. al.} 1998]{mazzotta98} 
   Mazzotta, P., Mazzitelli, G., Colafrancesco, S., Vittorio, N., 1998,
   A\&AS, 133, 403  
  \bibitem[\protect\citename{Meyer \& Meyer-Hofmeister} 1983]{meyer83}
    Meyer, F., Meyer-Hofmeister, E., 1983,
    A\&A, 121, 29
  \bibitem[\protect\citename{Papaloizou \& Bath} 1975]{papaloizou75}
    Papaloizu, J.\ C.\ B, Bath, G.\ T., 1975,
    MNRAS, 172, 339
  \bibitem[\protect\citename{Patterson }1979]{patterson79}Patterson~J.,
    1979, ApJ, 234, 978
  \bibitem[\protect\citename{Patterson et al.} 1980]{patterson80}
    Patterson, J., Branch, D., Chincarini, G., Robinson, E.\ L.,
    1980, ApJ, 240, L133
  \bibitem[\protect\citename{Pequignot, Petitjean \& Boisson} 1991]
   {pequignot91} 
   Pequignot, D., Petitjean, P., Boisson, C., 1991,
   A\&A, 251, 680
  \bibitem[\protect\citename{Press et al.} 1986]{numrec}
    Press, W.\ H., Teukolsky, S.\ A., Vetterling, W.\ T., Flannery, B.\ P.,
    1986, Numerical Recipes in Fortran, Cambridge Univ. Press
  \bibitem[\protect\citename{Ritter} 1988]{ritter88}
    Ritter, H., 1988, A\&A, 202, 93
  \bibitem[\protect\citename{Sarna} 1990]{sarna90}
    Sarna, M.\ J., 1990, A\&A, 239, 163
  \bibitem[\protect\citename{Shull \& Van Steenberg} 1982]{shull82}
    Shull, J.\ M., Van Steenberg, M., 1982, ApJS, 48,95
  \bibitem[\protect\citename{Skidmore et al. }2002]{skidmore02}
    Skidmore, W., O'Brien, K., Horne, K.\ D., 
    Gomer, R., Oke, J.\ B., Pearson, K.\ J., 
    2002, MNRAS, submitted
  \bibitem[\protect\citename{Stilley \& Callaway} 1970]{stilley70}
    Stilley, J.\ L., Callaway, J., 1970, ApJ, 160, 245
  \bibitem[\protect\citename{Thorstensen et al.} 1991]{thorstensen91}
    Thorstensen, J.\ R., Ringwald, F.\ A., Wada, R.\ A., Schmidt, G.\ A., 
    Norsworthy, J.\ E., 1991, AJ, 102, 272
  \bibitem[\protect\citename{van~Paradijs, Kraakman \& van~Amerongen} 1989]
    {paradijs89}van~Paradijs~J., Kraakman~H., van~Amerongen~S., 1989, A\&AS, 
    79, 205
  \bibitem[\protect\citename{Verner \& Ferland } 1996]{verner96}
    Verner, D.\ A., Ferland, G.\ J., 1996, ApJS, 103, 467
  \bibitem[\protect\citename{Verner et al.} 1996]{verner96b}
    Verner, D.\ A., Ferland, G.\ J., Korista, K.\ T., Yakovlev, D. G., 1996,
    ApJ, 465, 487
  \bibitem[\protect\citename{Voronov} 1997]{voronov97}
    Voronov, G.\ S., 1997, Atomic Data and Nuclear Data Tables, 65, 1
  \bibitem[\protect\citename{Warner }1995]{warner95}
    Warner, B., 1995, Cataclysmic Variable Stars, Cambridge Univ. Press
  \bibitem[\protect\citename{Welsh, Horne \& Oke }1993]
    {welsh93}Welsh~W.F., Horne~K., Oke~B., 1993, ApJ, 406, 229
  \bibitem[\protect\citename{Welsh, Horne \& Gomer }1995]{welsh95}
    Welsh, W.\ F., Horne, K.\ D., Gomer, R., 1995, MNRAS, 275, 649
 \bibitem[\protect\citename{Welsh, Horne \& Gomer }1998]{welsh98}
    Welsh, W.\ F., Horne, K.\ D., Gomer, R., 1998, MNRAS, 298, 285
  \bibitem[\protect\citename{Wynn, King \& Horne }1995]{wynn95}
    Wynn, G.\ A., King, A.\ R., Horne, K.\ D., 1995, in Buckley, D.\ A.\ H.,
    Warner,\ B., eds., Cape Workshop on Magnetic Cataclysmic Variables, 
    ASP Conf. Series, 85, 196, Astron. Soc. Pacific.,  San Francisco
  \bibitem[\protect\citename{Wynn, King \& Horne }1997]{wynn97}
    Wynn, G.\ A., King, A.\ R., Horne, K.\ D., 1997, MNRAS, 286, 436

\end{thebibliography}
\end{document}